\documentclass[twocolumn,showpacs,aps,prb,superscriptaddress,floatfix,amsmath,amssymb]{revtex4}
\usepackage{graphicx}% Include figure files
\usepackage{dcolumn}% Align table columns on decimal point
\usepackage{bm}% bold math
\usepackage{subfigure}

\begin{document}

\title{Structural transition and orbital glass physics in near-itinerant $\mathrm{CoV_{2}O_{4}}$.}

\author{D. Reig-i-Plessis}
\affiliation{Department of Physics and the Seitz Materials Research Laboratory, University of Illinois at Urbana-Champaign, Urbana, Illinois, USA, 61801}

\author{D. Casavant}
\affiliation{Department of Physics and the Seitz Materials Research Laboratory, University of Illinois at Urbana-Champaign, Urbana, Illinois, USA, 61801}

\author{V. O. Garlea}
\affiliation{Quantum Condensed Matter Division, Oak Ridge National Laboratory, Oak Ridge, Tennessee, 37831, USA}

\author{A. A. Aczel}
\affiliation{Quantum Condensed Matter Division, Oak Ridge National Laboratory, Oak Ridge, Tennessee, 37831, USA}

\author{M. Feygenson}
\affiliation{Chemical and Engineering Materials Division, Oak Ridge National Laboratory, Oak Ridge, Tennessee, 37831, USA}

\author{J. Neuefeind}
\affiliation{Chemical and Engineering Materials Division, Oak Ridge National Laboratory, Oak Ridge, Tennessee, 37831, USA}

\author{H. D. Zhou}
\affiliation{Department of Physics and Astronomy, University of Tennessee, Knoxville, TN, 37996, USA}
\affiliation{National High Magnetic
Field Laboratory, Florida State University, Tallahassee, FL
32306-4005, USA}

\author{S. E. Nagler}
\affiliation{Quantum Condensed Matter Division, Oak Ridge National Laboratory, Oak Ridge, Tennessee, 37831, USA}
\affiliation{Bredesen Center, University of Tennessee, Knoxville, TN, USA, 37996}

\author{G. J. MacDougall}
\email{gmacdoug@illinois.edu}
\affiliation{Department of Physics and the Seitz Materials Research Laboratory, University of Illinois at Urbana-Champaign, Urbana, Illinois, USA, 61801}

\date{\today}
\begin{abstract}

The ferrimagnetic spinel $\mathrm{CoV_2O_4}$ has been a topic of intense recent interest, both as a frustrated insulator with unquenched orbital degeneracy and as a near-itinerant magnet which can be driven metallic with moderate applied pressure. Here, we report on our recent neutron diffraction and inelastic scattering measurements on powders with minimal cation site disorder. Our main new result is the identification of a weak ($\frac{\Delta a}{a} \sim 10^{-4}$), first order structural phase transition at $T^*$ = 90 K, the same temperature where spin canting was seen in recent single crystal measurements. This transition is characterized by a short-range distortion of oxygen octahedral positions, and inelastic data further establish a weak $\Delta\sim 1.25 meV$ spin gap at low temperature. Together, these findings provide strong support for the local orbital picture and the existence of an orbital glass state at temperatures below $T^*$.

\end{abstract}

%PACS numbers {neutron diffraction for structures, spin arrangements in magnetically ordered materials, orbital charge or other orders, magnetic oxides, neutron inelastic
\pacs{61.05.fm,75.25.-j,75.25.Dk, 78.70.Nx}

\maketitle

\section{Introduction}

The spinel vanadates, $\mathrm{AV_2O_4}$, are interesting materials which lay at the heart of several distinct topics in condensed matter physics. In recent years, the focus has been geometric frustration, and the role played by the vanadium orbital angular momentum degrees of freedom in relieving the significant spin frustration inherent to the spinel structure\cite{lee10}. In insulating vanadates containing divalent $A$-site cations, the octahedrally coordinated $\mathrm{V^{3+}}$ comprise a pyrochlore sublattice with $S = 1$ spins and an orbital triplet degeneracy in the ideal cubic crystal field environment. When the $\mathrm{A^{2+}}$ cation is non-magnetic, this degeneracy is relieved at a Jahn-Teller structural transition from cubic-to-tetragonal global symmetry, followed by an antiferromagnetic spin order transition upon cooling\cite{mamiya97,ueda97,zhang06,nishiguchi02,onodo03,reehuis03, lee04,wheeler10,mun14}. In materials $\mathrm{MnV_2O_4}$\cite{adachi05, suzuki07_2, zhou07, garlea08, chung08, nii13, gleason14, nii12} and $\mathrm{FeV_2O_4}$\cite{katsufuji08,macdougall12,zhang12,nii12,kang12,macdougall14,zhang14}, the $A$-site also contains a spin, and $A-B$ superexchange interactions stabilize a form of collinear spin order, before orbital order transitions simultaneously distort the cubic structure and cant spins at lower temperatures. $\mathrm{CoV_2O_4}$ is a unique case. Though a collinear ferrimagnetic transition near $\mathrm{T_{N1}}$ = 150 K\cite{menyuk62,dwight64,kismarahardja11,huang12,koborinai15} has long been observed and there have been recent reports suggesting a weaker spin canting transition near $\mathrm{T_{N2}}$ = 90K\cite{koborinai15}, the material is reported to retain cubic global symmetry to lowest measured temperatures\cite{kismarahardja11}. This last observation is especially interesting, as it contradicts all known models for this class of materials. We discuss it in more detail below.

Enhanced spin-lattice coupling and associated phenomena resulting from orbital order have been reported on several occasions, and theory has predicted a number of different ordering patterns\cite{tsunetsugu03,motome04,tchernyshyov04,maitra07,perkins07,sarkar09,chern10,sarkar11}. Given the difficulty of accessing the orbital order parameter experimentally, conclusions have been argued indirectly, and debate has centered on the relative importance of exchange, spin-orbit coupling and various crystal field terms in the magnetic Hamiltonian. A secondary line of debate has focussed on whether localized orbital models are appropriate at all, or whether one should approach the problem from the itinerant limit for some materials\cite{rogers63,rogers64,blanco07,pardo08,kaur14}. Indeed, the spinel vanadates were initially studied in the context of localized-itinerant crossover physics, and are predicted to be metallic for V-V separations ($\mathrm{R_{VV}}$) less than the critical distance of $\mathrm{R_c}$ = 2.97\AA\cite{rogers63}. Though known materials are all semiconductors, those with $\mathrm{R_{VV} \sim R_c}$  (e.g. $\mathrm{MgV_2O_4}$, $\mathrm{ZnV_2O_4}$ and $\mathrm{CoV_2O_4}$) have enhanced electron-lattice coupling and transport properties best described by the hopping of polarons\cite{rogers63}. The activation energy of polarons is seen to decrease sharply (i.e. materials become more metallic) with decreasing $\mathrm{R_{VV}}$\cite{rogers63,blanco07} and increasing pressure\cite{kismarahardja11}, but increase with disorder due to cation inversion\cite{rogers63} or mixed valence on the spinel B-site\cite{rogers64}. $\mathrm{CoV_2O_4}$, which has the shortest $\mathrm{R_{VV}}$, has been driven metallic over a limited temperature range with an applied pressure of 6 GPa\cite{kismarahardja11}.

Recent studies have expounded on the role played by the proximate itinerant state. One study showed that, although materials $\mathrm{CdV_2O_4}$ and $\mathrm{MnV_2O_4}$ are well described by insulating models, the $Mn$ compound exhibits anomalously high compressibility which has been argued to signify a breakdown of the Mott-Hubbard picture\cite{blanco07,kismarahardja13}. Materials with shorter $\mathrm{R_{VV}}$ ($A \in \{Fe,Zn,Mg,Co\}$) may lay in an intermediate phase, characterized by partial delocalization, mobility of large polarons and short-range magnetic order\cite{blanco07}. These arguments seem to be backed by first principles calculations in $\mathrm{CoV_2O_4}$, which suggest the coexistence of localized and itinerant electrons\cite{kaur14}, and electronic structure calculations in $\mathrm{ZnV_2O_4}$, which explain the low-temperature transitions as a structural instability toward V-V dimerization, with no role for local orbitals\cite{pardo08}. The role of orbital order in near-itinerant vanadates was further challenged by studies of the doping series $\mathrm{Mn_{1-x}Co_xV_2O_4}$, where x-ray and heat capacity measurements reveal the suppression of the low temperature tetragonal phase transition above critical doping $x_c\sim0.8$\cite{kiswandhi11} and neutron scattering reveals the decoupling of this transition from spin canting\cite{ma15}.

In light of the above discussion, the case of $\mathrm{CoV_2O_4}$ is particularly intriguing. This material is a direct analogue to $\mathrm{MnV_2O_4}$, forming a near ideal spinel structure with $\mathrm{V^{3+}}$ on the pyrochlore lattice and spin-only S=3/2 $\mathrm{Co^{2+}}$ on the $A$-site. Its near critical V-V separation has made it the subject of several studies exploring the role of the proximate itinerancy in 3d magnet systems\cite{rogers63,rogers64,kismarahardja11,kiswandhi11,kismarahardja13,ma15}. Though it is well-established that there exists a magnetic transition to a near collinear ferrimagnetic ordered state  below $\mathrm{T_{N1}\sim 150 K}$\cite{menyuk62,rogers63,dwight64, kismarahardja11,huang12, kismarahardja13, koborinai15}, the experimental situation at lower temperatures is quite unclear. While one study of  $\mathrm{CoV_2O_4}$ powders reported a cusp in magnetization near T = 100 K and a sharp peak in heat capacity at T = 60 K\cite{huang12}, similar measurements of a single crystal sample revealed only a magnetic anomaly at T = 75 K and no signature in heat capacity at any temperature below $\mathrm{T_{N1}}$\cite{kismarahardja11}.  Neither study saw a deviation from cubic symmetry down to T = 10 K\cite{kismarahardja11,huang12}, which itself is surprising, as both orbital order\cite{tsunetsugu03,motome04,tchernyshyov04,maitra07,perkins07,sarkar09,chern10,sarkar11} and dimerization models\cite{khomskii05,pardo08} predict a symmetry lowering structural transition. Very recently, a neutron and strain study of crystals and powders enriched with excess cobalt reported a spin canting transition $\mathrm{T_{N2} = 90 K}$, which they associated with the onset of an orbital glassiness, and a separate structural transition of unknown origin near T = 40 K\cite{koborinai15}.

The large variation in the literature is likely a reflection of the combined weakness of observed effects and the enhanced sensitivity of this material to disorder. The differing heat capacity signatures in powders and crystals has been ascribed to the presence of random fields in crystals\cite{huang12}, as seen in $\mathrm{FeCr_2S_4}$\cite{fichtl05}, or could simply reflect the difficulty in detecting first order transitions in heat capacity using the time-relaxation technique, as suggested elsewhere by us\cite{macdougall11}. Weak inversion, and in particular cobalt on the spinel $B$-site, is known to affect both the temperature of transitions\cite{rogers63,koborinai15} and the degree of localization\cite{rogers63,rogers64,rogers66} in this system. In fact, early magnetization measurements noted a feature near 60 K, which seemed to disappear upon annealing and was attributed to the presence of $\mathrm{Co_2VO_4}$\cite{menyuk62}. Clearly, further measurements are needed before reaching final conclusions.

In this article, we shed light on the outstanding questions surrounding $\mathrm{CoV_2O_4}$ using a combination of neutron powder diffraction and inelastic scattering. Our diffraction results confirm the near cubic spinel structure with minimum cation inversion, and canted ferrimagnetic order below $\mathrm{T_N = 156 K}$. Significantly though, our analysis further reveals the presence of a weak, first-order structural phase transition at $\mathrm{T^{*}}$ = 90 K, the same temperature at which a spin canting transition is reported in Ref.~\onlinecite{koborinai15}. This transition is evident in the temperature evolution of both the lattice parameter, seen on long length-scales, and the oxygen atom positions, which has a short-range character. Our inelastic data show the success of local spin models in describing the magnon spectrum, and suggest the presence of a small (1.25 meV) spin gap. Together, these observations strongly suggest that this material is most appropriately described by local orbital models, and indicate an orbital ordering transition at $\mathrm{T^*}$. Proximate itinerancy is manifest primarily through a strong reduction of vanadium ordered moments, canting angle and the magnitude of the structural distortion.

\section{Materials Preparation and Experimental Details}

Powder samples of $\mathrm{CoV_2O_4}$ were prepared via solid state reaction at the National High Magnetic Field Lab in Tallahassee, FL, and extensive characterization data can be found on related samples in Refs.~\onlinecite{kismarahardja11} and \onlinecite{kismarahardja13}. As discussed above, these references report a near cubic structure to lowest temperatures and the formation of a net ordered moment below 150K, reflecting the ferrimagnetic spin order known from previous neutron measurements\cite{dwight64}. Neutron scattering data in the current report was obtained entirely using instruments at Oak Ridge National Laboratory (ORNL) in Oak Ridge, TN. Powder diffraction work was performed using the HB2A instrument at the High Flux Isotope Reactor (HFIR)\cite{hb2a}, followed by wide-angle time-of-flight (TOF) measurements using the Nanoscale-Ordered Materials Diffractometer (NOMAD) instrument at the Spallation Neutron Source (SNS)\cite{nomad}. HB2A measurements were performed using $\sim$3g of powder, placed in a vanadium can with helium exchange gas to ensure cooling. Patterns were measured independently with $\lambda = 1.538$\AA and $\lambda = 2.41$\AA~ neutrons from a germanium monochromator, and fit simultaneously. For NOMAD measurements, about 100 mg of sample was measured in 6 mm diameter vanadium cans for 20 minutes in an ILL Orange cryostat. NOMAD detectors were calibrated using scattering from diamond powder, and the instrument parameter file for the Rietveld refinements was obtained from the measurements of the NIST standard silicon powder. The structure factor S(Q)  was obtained by normalizing diffraction data against a solid V rod, and subtracting the background using the IDL routines developed for the NOMAD instrument. The pair-distribution function (PDF) was obtained by the Fourier transform of S(Q) with Q$_{max}$ = 31.5\AA$^{-1}$:

\begin{equation}
\label{eq:pdfformula} G(r)=\frac{2}{\pi}\int_{Q_{min}}^{Q_{max}}Q(S(Q)-1)\sin(Qr)dQ
\end{equation}

Rietveld refinements were performed using Fullprof\cite{fullprof}, and PDF fits were performed using the PDFgui software\cite{pdfgui}.

Inelastic neutron scattering (INS) measurements were performed using the hybrid spectrometer HYSPEC at the SNS.  HYSPEC is a highly versatile direct geometry spectrometer that combines TOF spectroscopy with the focusing Bragg optics\cite{hyspec}. The incident neutron beam is monochromated using a Fermi chopper with short, straight blades, and is then vertically focused by Bragg scattering onto the sample position by highly oriented pyrolytic graphite. HYSPEC employs 3He linear position sensitive tube detectors that are assembled into 20 sets of 8-packs that cover an angular range of 60$^\circ$ in the horizontal scattering plane and a vertical acceptance of 15$^\circ$.  For the INS measurements, the $\mathrm{CoV_2O_4}$ powder was loaded in an aluminium can and placed in a closed cycle refrigerator capable of reaching a base temperature of 5 K. Data was collected with incident energies Ei = 60 meV, 35 meV, 15 meV and 7.5 meV, and Fermi chopper frequencies of  300 Hz, 180 Hz, 180 Hz and 420 Hz, respectively.

\section{Neutron Scattering Results}

\subsection{Powder Diffraction}

\begin{figure}[t]
\begin{center}
\includegraphics[width=0.8\columnwidth]{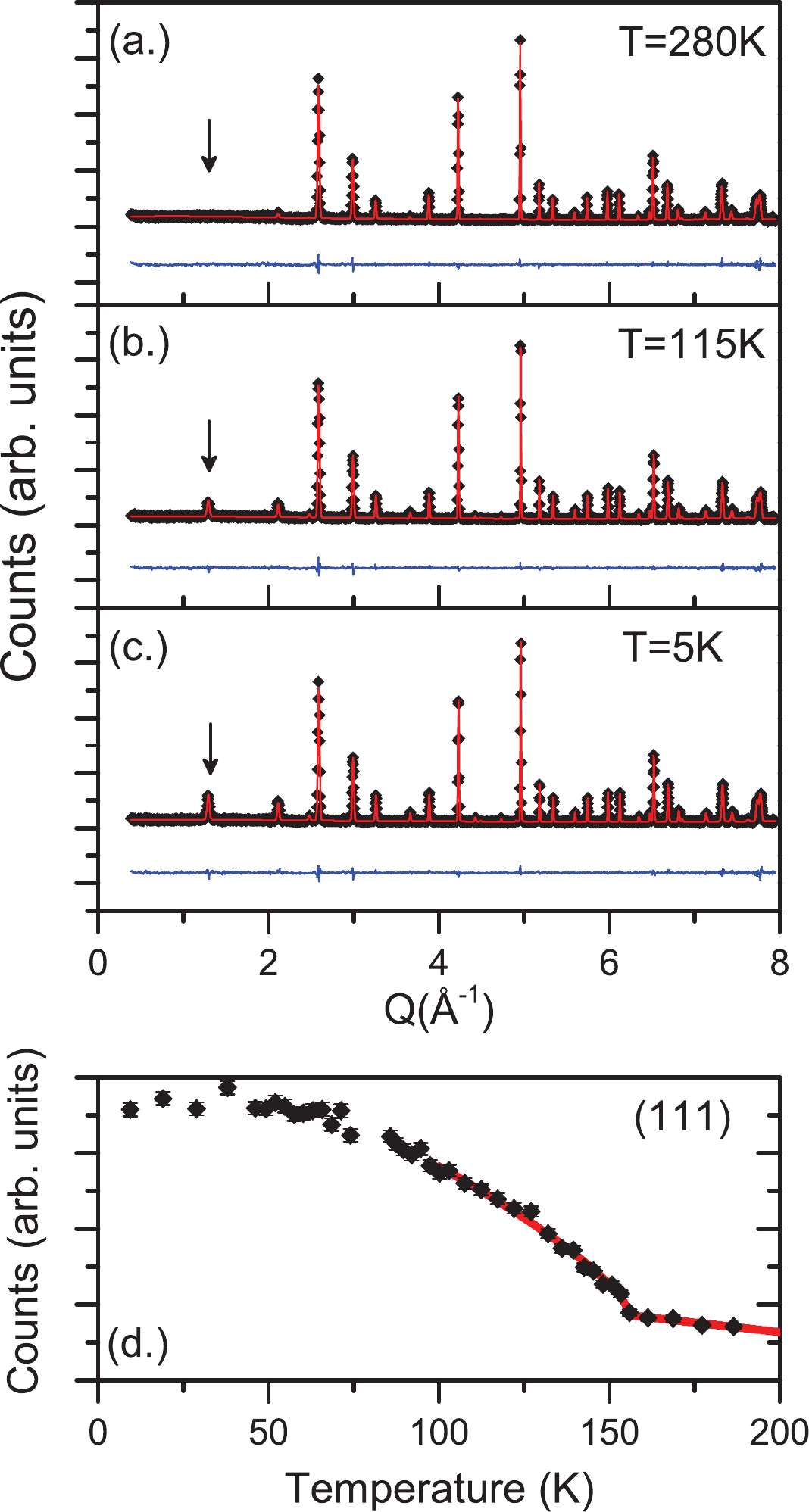}
\caption{ Powder diffraction patterns taken using the HB2A diffractometer with $\lambda = 1.54$\AA~ at temperatures T = 280 K (a.), 115 K (b.) and 5 K(c.). Here, black dots are data and solid lines are the results of Rietveld refinements described in the main text. Arrows denote the position of the cubic (111) Bragg reflection, which serves as an order parameter for the collinear ferrimagnetic state. A power law fit to the temperature variation of this peak (d.) gives $\mathrm{T_N = 156.1 \pm 0.5 K}$.}\label{fig:HB2a}
\end{center}
\end{figure}

Neutron powder diffraction measurements were first performed using the HB2A instrument with two neutron wavelengths, and main results are shown in Figure~\ref{fig:HB2a}. Figs.~\ref{fig:HB2a} (a.)-(c.) shows patterns from 1.54\AA~ measurements at T = 280 K, 115 K and 4 K, along with lines showing best-fit Rietveld refinements, performed using both wavelengths. Refinement of the pattern at T = 280 K (Fig.~\ref{fig:HB2a}(a)) confirms the near ideal spinel structure, with approximately 4$\%$ of the vanadium sites containing cobalt cations. Patterns at lower temperatures (Fig.~\ref{fig:HB2a}(b)-(c)) exhibit a visible increase in the intensity of several low-angle peaks associated with collinear ferrimagnetism, most prominently the cubic (111). Targeted measurement of this peak over a range of temperatures resulted in the data in Fig.~\ref{fig:HB2a}(d), and simple power-law fits yield a transition temperature of $\mathrm{T_N = 156.1 \pm 0.5 K}$ and a power $\mathrm{2\beta = 0.68 \pm 0.04}$. Consistent with other neutron powder diffraction studies\cite{dwight64,koborinai15}, we saw no visible intensity at the position of the cubic (200) Bragg peak associated with spin canting, yet refinements of the entire patterns to a model with $\mathrm{Co^{2+}}$ spins along the $c$-axis and $\mathrm{V^{3+}}$ spins allowed to cant along $<$110$>$ from the antiparallel direction implied a small canting angle at base temperature. Specifically, at T = 4 K, best fits give ordered moments $\mathrm{M_{Co} = 3.05(4) \mu_B}$, $\mathrm{M_{V,||} = 0.70(4) \mu_B}$, and $\mathrm{M_{V_\perp} = 0.06(4) \mu_B}$, where $\mathrm{M_{V,||}}$ and $\mathrm{M_{V,\perp}}$ denote ordered vanadium moments antiparallel and perpendicular to $\mathrm{M_{Co}}$, respectively. The value for $\mathrm{M_{Co}}$ is within error equal to the expected full-moment value for S=3/2 $\mathrm{Co^{2+}}$. The fitted values for $\mathrm{M_{V,||}}$ and $\mathrm{M_{V_\perp}}$ within our model imply a greatly reduced vanadium moment of $\mathrm{M_V = 0.70\mu_B}$, with a canting of $\mathrm{5 \pm 4 ^\circ}$. Though the fitted canting angle is not significantly different from zero, this result is consistent with the reports of canting in Ref.~\onlinecite{koborinai15} for Co-rich samples and Ref.~\onlinecite{ma15} for near itinerant members of the series $\mathrm{Mn_{2-x}Co_xV_2O_4}$. A complete listing of refinement parameters for each of the four measured temperatures is listed in Table~\ref{table:HB2a_params}.

\begin{table}[h]
\centering
\begin{tabular}{c c c c c c}
T (K) & a (\AA) & $x_{O^{2-}}$ & $M_{Co}$ ($\mu_B$) & $M_{V,||}$ ($\mu_B$) & $M_{V,\perp}$ ($\mu_B$)  \\
\hline
280 & 8.39976(8) & 0.23965(7) & - & - & - \\
115 & 8.39102(8) & 0.23971(7) & 2.43(4) & 0.60(4) & 0.04(4) \\
85 & 8.39007(7) & 0.23965 & 2.78(4) & 0.67(4) & 0.05(4) \\
4 & 8.38924(7) & 0.23964(7) & 3.05(4) & 0.70(4) & 0.06(4) \\
\hline
\end{tabular}
\caption{Parameters extracted from simultaneous Rietveld refinements of NPD patterns taken using $\lambda = 1.54$\AA~ and $2.41$\AA~ neutrons on the HB2A instrument at HFIR. Details about the fit model are provided in the main text.}
\label{table:HB2a_params}
\end{table}

\subsection{TOF Diffraction and Pair-Distribution Function Analysis}

\begin{figure}[tb]
\begin{center}
\includegraphics[width=0.8\columnwidth]{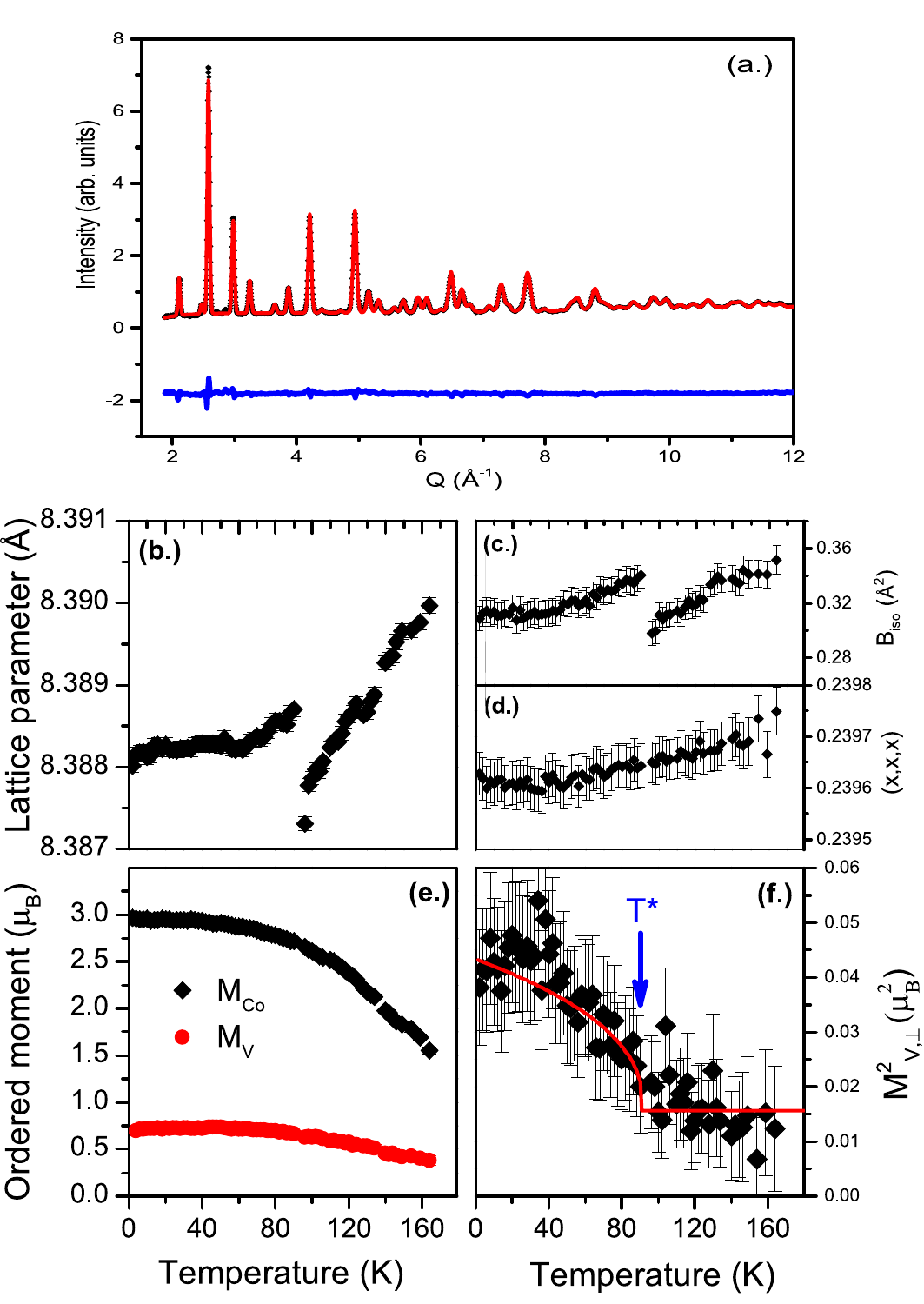}
\caption{Main results from a Rietveld refinement of diffraction data taken with the NOMAD instrument. TOF data were fit to the ideal cubic structure and the canted spin model described in the main text. Panel \textbf{(a.)} shows a typical pattern from data collected using a single NOMAD detector bank. Here, the solid line shows the best fit to the above model. The temperature dependence shows a jump in the fit lattice parameter (panel \textbf{(b.)}) and atomic displacement parameter for oxygens (panel \textbf{(c.)}) at the same temperature, $\mathrm{T^{*}}$ = 90 K. Neither the fractional position parameter for the oxygens (panel \textbf{(d.)}), nor the ordered moment sizes (panel \textbf{(e.)}) show a signature at $\mathrm{T^{*}}$. Panel \textbf{(f.)} is the square of the transverse $\mathrm{V^{3+}}$ ordered moment, equivalent to the cubic (200) Bragg peak, commonly thought to indicate spin canting. Solid line in this panel is a fit to a power-law temperature dependence, and implies $\mathrm{T^* = 92 K \pm 5 K}$.}\label{fig:NOMAD_FP}
\end{center}
\end{figure}

\begin{figure}[th]
\begin{center}
\includegraphics[width=0.8\columnwidth]{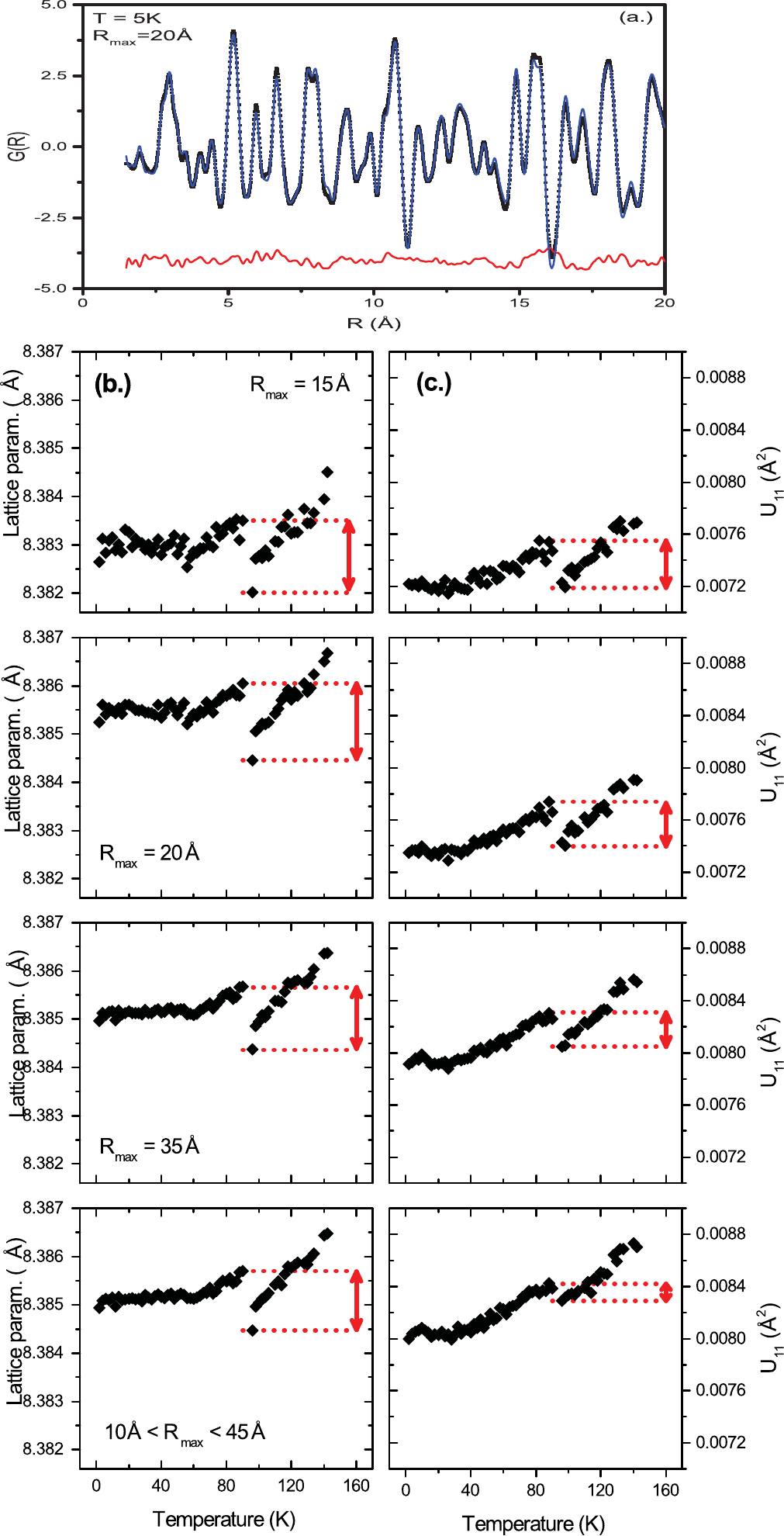}
\caption{Results of a PDF analysis of the NOMAD data, assuming a cubic spinel structure and neglecting scattering from spin order. Panel \textbf{(a.)} shows a typical fit at base temperature, demonstrating the ability of the model to describe the PDF patterns. Panels \textbf{(b.)} plot the refined lattice parameter as a function of temperature, as inferred from fits of real space data over several distinct data ranges. From top to bottom, data was fit using the range $\mathrm{1\AA~ < R < R_{max}}$, for $\mathrm{R_{max} = 15}$\AA, 20\AA, and 35\AA, whereas the bottom panel used $\mathrm{10\AA~ < R < 45\AA}$. Panels \textbf{(c.)} show equivalent plots of the oxygen ADP. Both sets of panels reveal a discontinuity at $\mathrm{T^* = 90 K}$, but exhibit a different dependence on fit range, as discussed in the main text. }\label{fig:NOMAD_PDF}
\end{center}
\end{figure}

To search for local distortions of oxygen octahedra or other signatures of orbital glassiness, we supplemented the above data with diffraction measurements using the NOMAD instrument, which has a significantly higher count rate and detector coverage more suitable for a PDF analysis. Data were analyzed using both Rietveld refinement in reciprocal space and with real-space PDF analysis, described above, and the main results are shown in Figs.~\ref{fig:NOMAD_FP} and \ref{fig:NOMAD_PDF}, respectively. Rietveld refinements used the same model and gave results largely consistent with conclusions from the HB2A analysis. Specifically, it was confirmed that the structure is very close to an ideal cubic spinel at all temperatures, and that spins order into the canted ferrimagnetic state described above. A representative pattern and associated fit line are shown in Fig.~\ref{fig:NOMAD_FP}(a.). Additionally, the sizeable increase in signal-to-noise on the NOMAD instrument brings to light several interesting new features. Despite the success of the cubic spinel model across the full temperature range, Fig.~\ref{fig:NOMAD_FP}(b.) shows a clear, discontinuous jump of $\mathrm{\Delta a = 0.0014}$\AA~ in the refined lattice parameter at a critical temperature $\mathrm{T^{*}}$ = 90 K. The magnitude of this jump is within the error bars of previous x-ray measurements\cite{kismarahardja11}, and this observation is consistent with previous literature. Moreover, though the fractional position parameter of the oxygen in the $\mathrm{Fd\bar{3}m}$ space group (Fig.~\ref{fig:NOMAD_FP}(c.)) shows a smooth and continuous variation with temperature, the atomic displacement parameter (ADP), $\mathrm{B_{iso}}$, which reflects the root mean square displacement of oxygen atoms from equilibrium, shows a jump similar to lattice parameter at the exact same $\mathrm{T^*}$. Together, these observations strongly suggest the existence of an unresolved, first-order structural phase transition at the limits of instrument resolution.

The existence of a structural transition at $\mathrm{T^{*}}$ = 90 K is particularly important in light of previous polarized neutron scattering work on single crystals, which has established that a spin-canting transition exists at the exact same temperature\cite{koborinai15}. Our own diffraction data is also not inconsistent with such a conclusion. In Table~\ref{table:NOMAD_FP_params}, we show refined magnetic moments at selected temperatures assuming the canted state described above\footnote{Parameters for all temperatures can be found in the Supplementary Materials.}.  One can see a smooth development of ferrimagnetism below $\mathrm{T_N}$, with moment sizes (Fig.~\ref{fig:NOMAD_FP}(e.)) equal within error to HB2A fits and an implied canting angle for vanadium spins of $\mathrm{15 \pm 5 ^\circ}$ at base temperature. This canting angle is statistically significant, though larger than the values in Table~\ref{table:HB2a_params} by a factor of 3-4; this reflects the difficulty of reliably determining this quantity from powder refinements alone. This canting angle is also a factor of 2 larger than implied by single crystal measurements in Ref.~\onlinecite{koborinai15}, but notably equal to results from powder refinements in the same study. In Fig.~\ref{fig:NOMAD_FP}(f.), we show the temperature variation of $\mathrm{M_{V_\perp}^2}$, which is proportional to the cubic (200) Bragg peak intensity, used in Ref~\onlinecite{koborinai15} to detect the spin canting transition in single crystals of this material. Though the magnitude of $\mathrm{M_{V_\perp}}$ (and thus canting angle) seems to grow with decreasing temperature, statistical error bars in our data set are too large to discern a clear kink at $\mathrm{T = T^{*}}$. We do note, however, that the total increase of $\mathrm{M_{V_\perp}^2}$ over the measured temperature range is the same as seen in Ref.~\onlinecite{koborinai15}, and if one assumes a transition, fits to a power-law temperature dependence yield  $\mathrm{T^* = 92 K \pm 5 K}$. Thus, we conclude that our data is consistent with the previously reported spin canting transition at $\mathrm{T^*}$\cite{koborinai15}. In light of this, the coincident structural transition seen in the current data set strongly suggests that this temperature can be associated with the ordering of $\mathrm{V^{3+}}$ orbitals.

\begin{table}[b]
\centering
\begin{tabular}{c c c c c c}
T (K) & a (\AA) & $x_{O^{2-}}$ & $M_{Co}$ ($\mu_B$) & $M_{V,||}$ ($\mu_B$) & $M_{V,\perp}$ ($\mu_B$) \\
\hline
144 & 8.3909(2) & 0.23968(5) & 2.02(4) & 0.53(5) & 0 \\
120 & 8.3902(2) & 0.23968(5) & 2.50(4) & 0.62(4) & 0.06(6) \\
100 & 8.3894(2) & 0.23966(5) & 2.74(3) & 0.68(4) & 0.08(6) \\
80 & 8.3907(2) & 0.23966(5) & 2.92(3) & 0.73(4) & 0.13(6) \\
60 & 8.3903(2) & 0.23963(5) & 3.01(3) & 0.73(3) & 0.17(6) \\
40 & 8.3905(2) & 0.23962(5) & 3.07(3) & 0.73(3) & 0.20(5) \\
20 & 8.3905(2) & 0.23962(5) & 3.09(3) & 0.73(3) & 0.21(5) \\
6 & 8.3905(2) & 0.23961(5) & 3.09(3) & 0.73(3) & 0.19(6) \\
\hline
\end{tabular}
\caption{Fit parameters at selected temperature, extracted from Rietveld refinements of NPD patterns taken using the NOMAD instrument at the SNS. Details about the fit model are provided in the main text, and further parameters for the entire temperature range can be found in the Supplemental Materials.}
\label{table:NOMAD_FP_params}
\end{table}

The same data set was subject to a PDF analysis, using the transform given by Eq.~\ref{eq:pdfformula}, and assuming a cubic spinel structure. The main results, shown in Fig.~\ref{fig:NOMAD_PDF}, tell a similar story to the one above. The cubic spinel structure describes the data well at all temperatures (see, e.g., Fig.~\ref{fig:NOMAD_PDF}(a.)), and there is a distinct jump in the inferred lattice parameter and the atomic displacement parameter of oxygen atoms at $\mathrm{T^* = 90 K}$. Fits were performed over a number of increasingly shortened ranges in real-space: 1.5\AA$\mathrm{ < R < R_{max}}$, with $\mathrm{R_{max}}$ varying from 30\AA~ to 15\AA. Additionally, one fit was performed using a range 10\AA$\mathrm{ < R < 45}$\AA, to explicitly suppress the effect of local correlations. As shown in Fig.~\ref{fig:NOMAD_PDF}(b.), the magnitude of the jump in lattice parameter at $\mathrm{T^*}$ was found to be largely independent of fit range, consistent with expectations for a true long-range effect. In contrast, the displacement of the oxygens decreases monotonically with increasing $\mathrm{R_{max}}$ and is strongly suppressed for the fit where local correlations are neglected. This suggests that the atomic displacement parameter of oxygens (Fig.~\ref{fig:NOMAD_PDF}(c.)) has an overwhelmingly local character. As this parameter represents the root mean square positions of oxygens forming the octahedra surrounding $\mathrm{V^{3+}}$ octahedra, this observation might suggest the onset of orbital glassiness, as previously conjectured\cite{koborinai15}. As with the Rietveld analysis, we were able to fit the PDF data assuming a number of different structural distortions at low temperature, but none showed a significant improvement when compared to the cubic spinel structure. The weakness of the current effect does not allow us to comment further.

\subsection{Inelastic Neutron Scattering}

\begin{figure}[t]
\begin{center}
\includegraphics[width=\columnwidth]{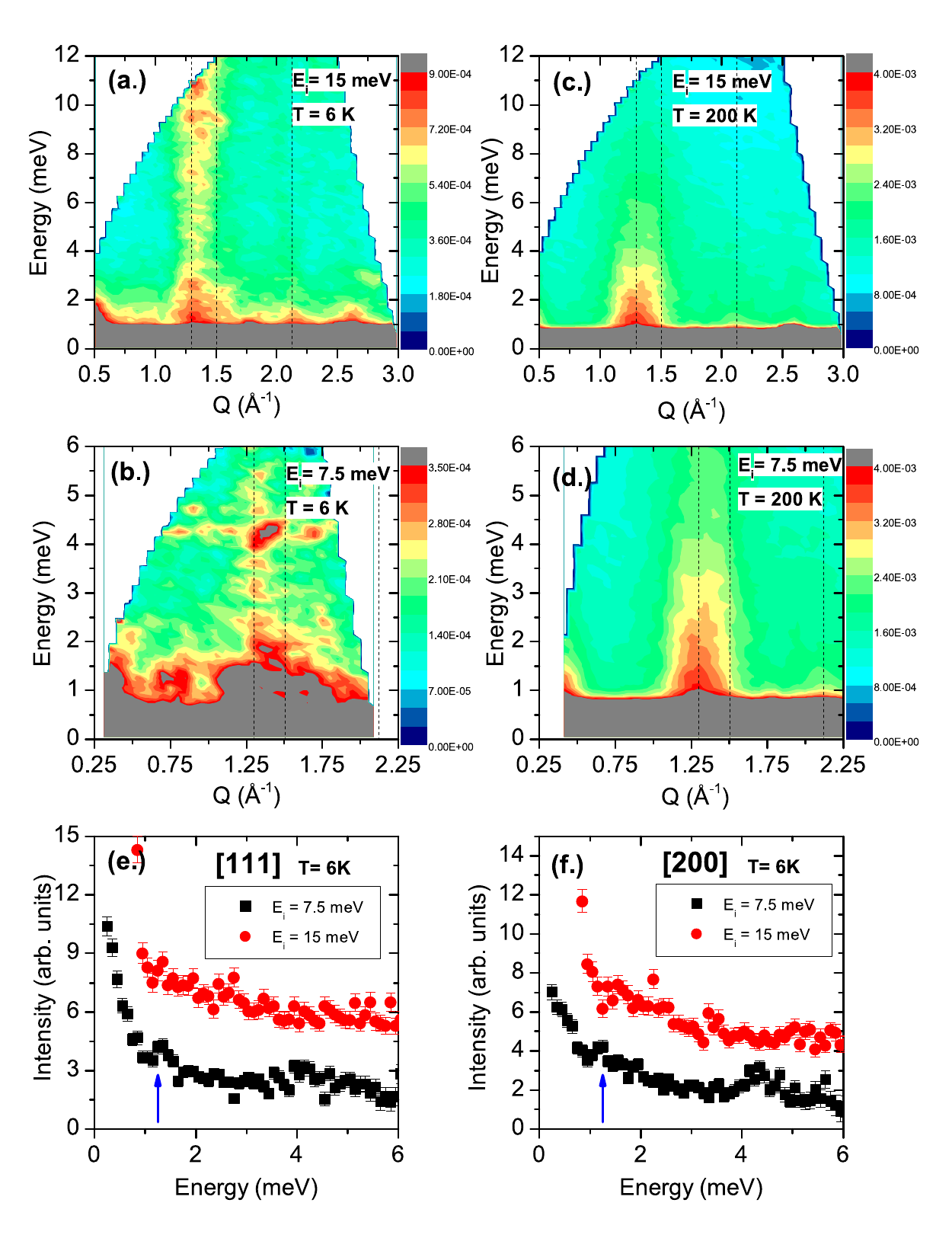}
\caption{Powder inelastic scattering data, taken using the HYSPEC instrument at the SNS. Panels (a.) and (b.) show data taken at base temperature, T= 6 K, with $\mathrm{E_i = 15 meV}$ and $\mathrm{E_i = 7.5 meV}$, respectively. The most obvious feature is the highly dispersive excitation, nearly touching the elastic line at the position of the (111) magnetic Bragg peak, and weaker bands are seen about (200) and (220). Vertical dotted lines denote the locations of the three aforementioned Bragg locations. Equivalent measurements taken at T = 200 K (panels (c.) and (d.)) show only correlated paramagnetic behavior, confirming that the above dispersive bands are magnetic in origin.  Constant-Q cuts at the positions of the (111) and (200) (panels (e.) and (f.)) reveal a small bump, consistent with a spin gap of $\Delta$ = 1.25 meV in this material. }\label{fig:HYSPEC}
\end{center}
\end{figure}

\begin{figure}[t]
\begin{center}
\includegraphics[width=\columnwidth]{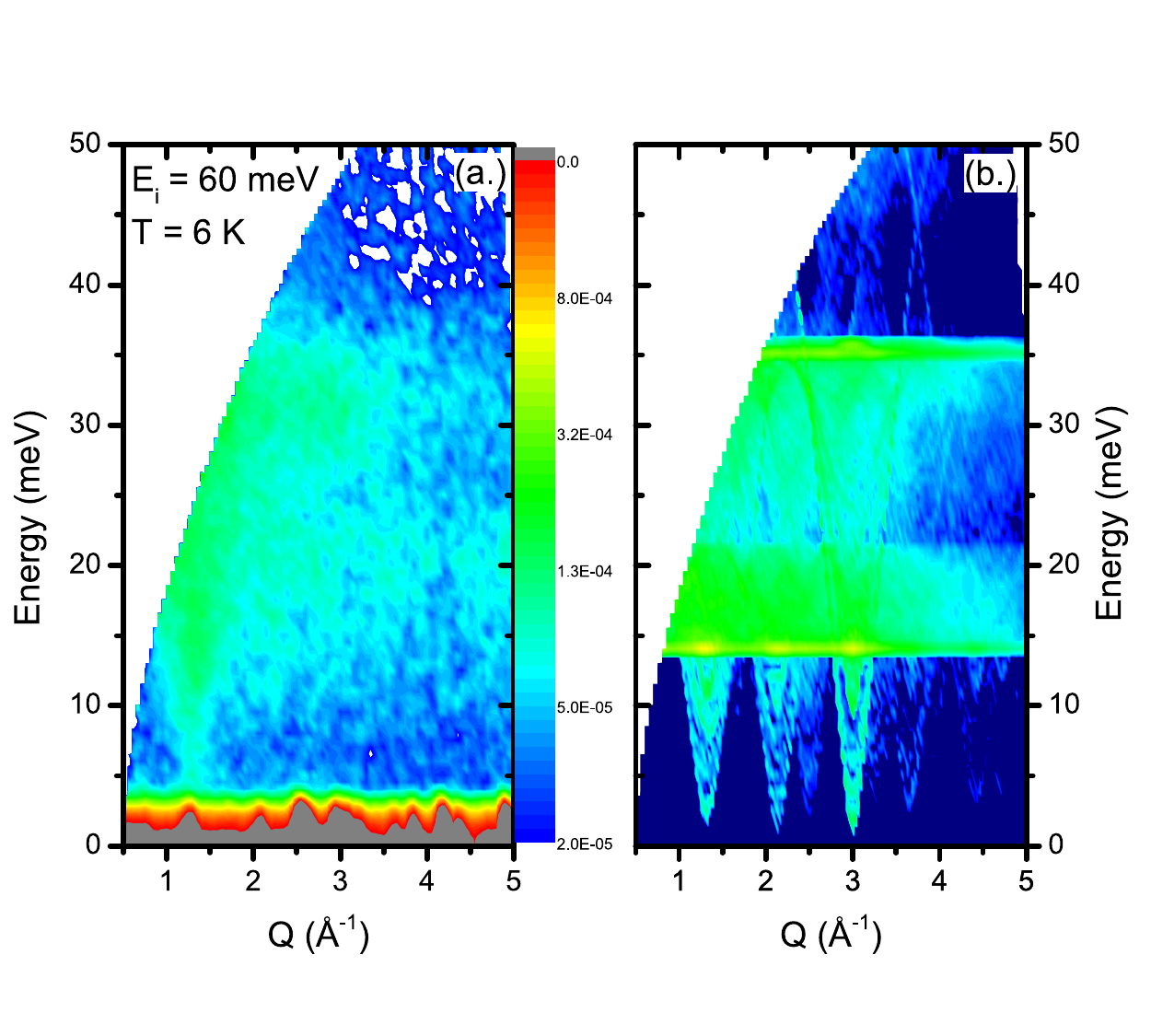}
\caption{Plots of measured (left) and simulated (right) neutron inelastic scattering data with $E_i = 60 meV$, plotted on a logarithmic scale. The simulated scattering data is the result of a spin-wave analysis, using parameters described in the main text and using spherical approximations for the local form factors of $Co^{2+}$ and $V^{3+}$ moments. Effects of instrument resolution have not been included.} \label{fig:SW}
\end{center}
\end{figure}

Further support for an orbital glass picture comes from inelastic neutron measurements on the HYSPEC spectrometer, summarized in Figs.~\ref{fig:HYSPEC} and \ref{fig:SW}. Figs.~\ref{fig:HYSPEC}(a.) and (b.) show the low-energy magnon spectra at T = 6 K, taken with $\mathrm{E_i}$ = 15 meV and 7.5 meV neutrons, respectively. Immediately apparent is the existence of a highly-dispersive, bright excitation mode emerging out of the position $\mathrm{Q = 1.3}$\AA$^{-1}$, which is associated with the cubic (111) Bragg peak. Closer inspection reveals weaker excitations emerging from $\mathrm{Q = 1.5}$\AA$^{-1}$ and $\mathrm{Q = 2.1}$\AA$^{-1}$, the location of the (200) and (220) peaks, respectively. Figs.~\ref{fig:HYSPEC}(c.) and (d.) confirm that these excitations give way to a correlated paramagnetic phase at temperatures T $\mathrm{> T_N}$.  The flat band of scattering in Fig.~\ref{fig:HYSPEC}(b.) is seen only with $\mathrm{E_i}$ = 7.5 meV, but not with $\mathrm{E_i}$ = 15 meV (Fig.~\ref{fig:HYSPEC}(a.)), 35 meV, or 60 meV (Fig.~\ref{fig:SW}(a.)). For this reason, we conclude it is likely spurious, though its origin is unknown.

Plots of scattering intensity versus energy transfer reveal a small peak in scattering at $\mathrm{\Delta\sim 1.25 meV}$ above both magnetic Bragg peaks, which is suggestive of a spin gap. Though we are unable to confirm the complete absence of scattering at lower energies due to experimental constraints, the magnitude of this gap is nearly a factor of 2 smaller than reported for $\mathrm{Mn_{2-x}Co_xV_2O_4}$ for $\mathrm{Co^{2+}}$ concentrations up to $x$ = 0.8\cite{ma15}. This is consistent with expectations for an orbital order state with strong renormalization due to proximate itinerancy.

In Fig.~\ref{fig:SW} we show a comparison of inelastic data taken with $\mathrm{E_i = 60 meV}$, and simulations based on a semiclassical spinwave Hamiltonian:

\begin{eqnarray}
 \mathcal{H} =  \sum\limits_{< i,j >} J_{i,j} \mathbf{S_i} \cdot \mathbf{S_j} +  \sum\limits_i D_i (\mathbf{S_i} \cdot \mathbf{\hat{n}_i})^2,
\end{eqnarray}

\noindent along with the parameters in Table~\ref{table:SW_params}. Here,the sums are over all spins, including both cation sites, and interactions are truncated beyond nearest neighbors for the $\mathrm{Co^{2+}}$ sites (nearest neighbor Co-V interactions) and next-nearest neighbors for the $\mathrm{V^{3+}}$ sites (nearest neighbor V-V and nearest neighbor Co-V interactions only). We have used $\mathrm{J_{AB}}$ to denote the exchange interaction between A-site cobalt and B-site vanadium spins, while $\mathrm{J_{BB}}$ ($\mathrm{J_{BB}^{'}}$) describes interactions between nearest-neighbor vanadium spins with the same (different) $c$-axis positional coordinates. Single-ion anisotropy was set to zero for cobalt ($\mathrm{D_A}$), consistent with the near-cubic structural symmetry, and fixed along local $<$111$>$ directions for the vanadium sites ($\mathrm{D_B}$). A similar analysis was successfully used by us to describe spin excitations in $\mathrm{FeV_2O_4}$\cite{macdougall14} and in Ref.~\onlinecite{ma15} to describe the $\mathrm{Mn_{1-x}Co_{x}V_2O_4}$ family of compounds.

Notably, the material parameters listed in Table~\ref{table:SW_params} and used to create Fig.~\ref{fig:SW}(b.) were \textit{not} extracted from fits to the inelastic data (Fig.~\ref{fig:SW}(a.)), but rather were extrapolated from parameters reported for single-crystalline $\mathrm{Mn_{0.4}Co_{0.6}V_2O_4}$\cite{ma15}, and adjusted to account for N$\mathrm{\acute{e}}$el temperature, canting angle and spin gap. It should be pointed then out that the Heisenberg exchange parameters in $\mathrm{Mn_{0.4}Co_{0.6}V_2O_4}$ are hugely anisotropic, as interactions between V-V pairs in that material are strongly modified by the existence of orbital order and the associated structural distortion. Specifically, orbital order results in $\mathrm{J_{BB}}$ and $\mathrm{J^{'}_{BB}}$ that differ significantly not only in magnitude, but also sign. Though the lattice structure in $\mathrm{CoV_2O_4}$ is more isotropic, we found we were unable to satisfactorily describe the inelastic data in Fig.~\ref{fig:SW} without retaining the anisotropy in the Heisenberg exchange parameters, and specifically the difference in sign between $\mathrm{J_{BB}}$ and $\mathrm{J^{'}_{BB}}$. The significance of this observation is discussed in more detail below.

Even without fitting, this basic spinwave model is able to account for several specific features of the $\mathrm{CoV_2O_4}$ data, including the minima at $1.3$~\AA$^{-1}$ and $1.5$~\AA$^{-1}$, the ridge of intense scattering between 14 meV and 20 meV, and the second ridge of scattering near 37 meV. Arguably, the largest failure of the model is an overestimation of scattering at higher Q, including the prediction of an intense mode near $Q = 3.0$\AA$^{-1}$. This observation indicates a deviation in $\mathrm{CoV_2O_4}$ from the assumed spherical approximation for the local magnetic form factor, and in fact may indicate that the atomic orbitals are distributed over larger distances in real space, as one might expect for a material approaching an itinerant crossover.

%\footnote{$J_{ab}$ was scaled to account for increased $T_N$. $D_a$ was set to zero to account for near cubic structure. $D_b$ was then adjusted to reproduce the observed gap, and $J_{bb}$+$J^{'}_{bb}$ were adjusted to account for difference in canting angle. It has been verified that equally good simulations can be obtained by a near-zero gap, through the same process.}
%jaa=0
%jab=1.8*5/3=3
%jbb=8
%jbbp=-4.59
%Da=0
%Db=-0.39
%cantangle=5deg
%ei=60meV

\begin{table}[b]
\centering
\begin{tabular}{c c c c c c }
$E_i$ & $J_{AB}$ & $J_{BB}$ & $J^{'}_{BB}$ & $D_A$ & $D_B$ \\
\hline
60 & 3 & 8 & -4.59 & 0 & -0.39 \\
\hline
\end{tabular}
\caption{Parameters used to produce simulated neutron scattering spectrum in Fig.~\ref{fig:SW}(b). All values are given in units of meV.}
\label{table:SW_params}
\end{table}

\section{Discussion and Conclusions}

Collectively, the neutron scattering results presented here draw a clear and consistent picture for the physics underlying $\mathrm{CoV_2O_4}$. The most striking new result is the first order structural phase transition at $T^*$ = 90K, which by itself addresses the largest outstanding question in the study of this material- namely the unexpected preservation of the high symmetry cubic phase to lowest temperatures. Here, we see that the predicted symmetry lowering transition is not absent, but has merely been suppressed in magnitude to $\frac{\Delta a}{a}\sim 10^{-4}$, presumably another effect of the near-itinerancy. The subtlety of the transition and our inability to resolve details of the low symmetry structure does not allow us to strictly determine whether it is driven by orbital order\cite{tsunetsugu03,motome04,tchernyshyov04,maitra07,perkins07,sarkar09,chern10,sarkar11} or an instability towards homopolar bond formation\cite{khomskii05,pardo08}. If one takes into account, however, previous reports of spin canting at the same temperature\cite{koborinai15} and the small spin gap seen in powder inelastic data at base temperature, the experimental situation surrounding $\mathrm{CoV_2O_4}$ is strongly reminiscent of other ferrimagnetic spinels more unambiguously described by the Mott insulating picture\cite{adachi05, suzuki07_2, zhou07, garlea08, chung08, nii13, gleason14, katsufuji08,macdougall12,zhang12,nii12,kang12,macdougall14,zhang14}. Thus, it seems most appropriate to describe the 90 K transition in $\mathrm{CoV_2O_4}$ as an ordering transition for localized $V^{3+}$ orbitals. In further support of this idea, we note that our successful description of inelastic scattering invoked a localized spin Hamiltonian, and used parameters which imply a strong anisotropy in Heseinberg exchange parameters. This magnetic anisotropy echoes previous reports of an anisotropic compression in single crystals\cite{kismarahardja13}. Without a much larger tetragonal distortion than what we are reporting, the most natural way of explaining these experimental signatures is though an emergent electronic anisotropy, such as one expects in an orbitally ordered state.

On a final note, we point out that our data allows us to rule out any additional structural or spin transitions below $\mathrm{T^{*}}$, within the sensitivity of our probe. This is in direct contrast to reports of transitions at 40 K\cite{koborinai15}, 60 K\cite{huang12} or 75 K\cite{kismarahardja11}, seen in both powders and single crystals. We believe the variation on this point in the literature is largely a result of the difficulty in preparing pure samples of this material and its peculiar sensitivity to disorder. It is known that the most frequent impurity phase in the preparation of $\mathrm{CoV_2O_4}$ is $\mathrm{V_2O_3}$, which is strongly correlated and has transition temperatures which are sensitive to level of defects\cite{mcwhan73}. On the other hand, it is also known that the presence of excess cobalt on the vanadium site strongly affects the degree of localization in this material\cite{rogers63,rogers64,rogers66}, drives changes in N$\mathrm{\acute{e}}$el temperature\cite{rogers63}, and has led to at least one report of a transition between 45 K and 70 K which disappears upon annealing\cite{menyuk62}. Moreover, it has been suggested that enhanced sensitivity to disorder as materials approach the itinerant crossover affects both the level of localization\cite{blanco07} and might further explain the many reports of glassy correlations and dynamics\cite{blanco07,huang12,kismarahardja11,kismarahardja13,koborinai15}. X-ray and neutron diffraction measurements on our samples reveal no detectable impurity phases and a cation inversion ($x\sim0.04$) that is among the lowest in the literature, which increases confidence in our results. A systematic multi-sample study will probably be required, however, to disentangle complicated disorder effects in this material and discern the true number of transitions.

In summary, we have presented strong evidence for a structural phase transition in $\mathrm{CoV_2O_4}$ at a temperature of $\mathrm{T^* = 90 K}$. This transition, coupled with previous reports of spin canting and the presence of a spin gap, lead us to believe that this material is best described by a local orbital ordering picture, though it is certainly possible that ferrimagnetic bond formation is playing a role. We suggest that the main effects of the nearby itinerant state are the strong reduction of the magnitude of the structural distortion, the reduction of the vanadium ordered moment and canting angle associated with the orbital order, and the modification of the magnetic form factor, indicative of a near delocalized spin distribution. Confirmation of these interpretations could come from future pressure experiments, where the known insulator-metal transition should completely suppress the reported structural transition and spin gap.

\section{Acknowledgements}

This work was sponsored by the National Science Foundation, under grant number DMR-1455264 (GJM, DR, DC). The work in NHMFL (HDZ) is supported by NSF-DMR-1157490 and the State of Florida. Research at Oak Ridge National Laboratory's High Flux Isotope Reactor and Spallation Neutron Source was sponsored by the Scientific User Facilities Division, Office of Basic Energy Sciences, U. S. Department of Energy. Authors further acknowledge useful conversations with S. Billinge, M. Matsuda and S.E. Dissanayake.

%\bibliography{./spinels_fullauthors}

\begin{thebibliography}{53}
\expandafter\ifx\csname natexlab\endcsname\relax\def\natexlab#1{#1}\fi
\expandafter\ifx\csname bibnamefont\endcsname\relax
  \def\bibnamefont#1{#1}\fi
\expandafter\ifx\csname bibfnamefont\endcsname\relax
  \def\bibfnamefont#1{#1}\fi
\expandafter\ifx\csname citenamefont\endcsname\relax
  \def\citenamefont#1{#1}\fi
\expandafter\ifx\csname url\endcsname\relax
  \def\url#1{\texttt{#1}}\fi
\expandafter\ifx\csname urlprefix\endcsname\relax\def\urlprefix{URL }\fi
\providecommand{\bibinfo}[2]{#2}
\providecommand{\eprint}[2][]{\url{#2}}

\bibitem[{\citenamefont{Lee et~al.}(2010)}]{lee10}
\bibinfo{author}{\bibfnamefont{S.-H.} \bibnamefont{Lee}} \bibnamefont{et~al.},
  \bibinfo{journal}{J.\ Phys.\ Soc.\ Jap.} \textbf{\bibinfo{volume}{79}},
  \bibinfo{pages}{011004} (\bibinfo{year}{2010}).

\bibitem[{\citenamefont{Mamiya et~al.}(1997)\citenamefont{Mamiya, Onoda,
  Furubayashi, Tang, and Nakatani}}]{mamiya97}
\bibinfo{author}{\bibfnamefont{H.}~\bibnamefont{Mamiya}},
  \bibinfo{author}{\bibfnamefont{M.}~\bibnamefont{Onoda}},
  \bibinfo{author}{\bibfnamefont{T.}~\bibnamefont{Furubayashi}},
  \bibinfo{author}{\bibfnamefont{J.}~\bibnamefont{Tang}}, \bibnamefont{and}
  \bibinfo{author}{\bibfnamefont{I.}~\bibnamefont{Nakatani}},
  \bibinfo{journal}{J.\ Appl.\ Phys.} \textbf{\bibinfo{volume}{81}},
  \bibinfo{pages}{5289} (\bibinfo{year}{1997}).

\bibitem[{\citenamefont{Ueda et~al.}(1997)\citenamefont{Ueda, Fujiwara, and
  Yasuoka}}]{ueda97}
\bibinfo{author}{\bibfnamefont{Y.}~\bibnamefont{Ueda}},
  \bibinfo{author}{\bibfnamefont{N.}~\bibnamefont{Fujiwara}}, \bibnamefont{and}
  \bibinfo{author}{\bibfnamefont{H.}~\bibnamefont{Yasuoka}},
  \bibinfo{journal}{J.\ Phys.\ Soc.\ Japan} \textbf{\bibinfo{volume}{66}},
  \bibinfo{pages}{778} (\bibinfo{year}{1997}).

\bibitem[{\citenamefont{Zhang et~al.}(2006)\citenamefont{Zhang, Louca,
  Visinoiu, Lee, Thompson, Proffen, Llobet, Qiu, Park, and Ueda}}]{zhang06}
\bibinfo{author}{\bibfnamefont{Z.}~\bibnamefont{Zhang}},
  \bibinfo{author}{\bibfnamefont{D.}~\bibnamefont{Louca}},
  \bibinfo{author}{\bibfnamefont{A.}~\bibnamefont{Visinoiu}},
  \bibinfo{author}{\bibfnamefont{S.-H.} \bibnamefont{Lee}},
  \bibinfo{author}{\bibfnamefont{J.~D.} \bibnamefont{Thompson}},
  \bibinfo{author}{\bibfnamefont{T.}~\bibnamefont{Proffen}},
  \bibinfo{author}{\bibfnamefont{A.}~\bibnamefont{Llobet}},
  \bibinfo{author}{\bibfnamefont{Y.}~\bibnamefont{Qiu}},
  \bibinfo{author}{\bibfnamefont{S.}~\bibnamefont{Park}}, \bibnamefont{and}
  \bibinfo{author}{\bibfnamefont{Y.}~\bibnamefont{Ueda}},
  \bibinfo{journal}{Phys.\ Rev.\ B} \textbf{\bibinfo{volume}{74}},
  \bibinfo{pages}{014108} (\bibinfo{year}{2006}).

\bibitem[{\citenamefont{Nishiguchi and Onoda}(2002)}]{nishiguchi02}
\bibinfo{author}{\bibfnamefont{N.}~\bibnamefont{Nishiguchi}} \bibnamefont{and}
  \bibinfo{author}{\bibfnamefont{M.}~\bibnamefont{Onoda}},
  \bibinfo{journal}{J.\ Phys.:\ Condens.\ Matter}
  \textbf{\bibinfo{volume}{14}}, \bibinfo{pages}{L551} (\bibinfo{year}{2002}).

\bibitem[{\citenamefont{Onoda and Hasegawa}(2003)}]{onodo03}
\bibinfo{author}{\bibfnamefont{M.}~\bibnamefont{Onoda}} \bibnamefont{and}
  \bibinfo{author}{\bibfnamefont{J.}~\bibnamefont{Hasegawa}},
  \bibinfo{journal}{J.\ Phys.: Condens.\ Matter} \textbf{\bibinfo{volume}{15}},
  \bibinfo{pages}{L95} (\bibinfo{year}{2003}).

\bibitem[{\citenamefont{Reehuis et~al.}(2003)\citenamefont{Reehuis, Krimmel,
  Buttgen, Loidl, and Prokofiev}}]{reehuis03}
\bibinfo{author}{\bibfnamefont{M.}~\bibnamefont{Reehuis}},
  \bibinfo{author}{\bibfnamefont{A.}~\bibnamefont{Krimmel}},
  \bibinfo{author}{\bibfnamefont{N.}~\bibnamefont{Buttgen}},
  \bibinfo{author}{\bibfnamefont{A.}~\bibnamefont{Loidl}}, \bibnamefont{and}
  \bibinfo{author}{\bibfnamefont{A.}~\bibnamefont{Prokofiev}},
  \bibinfo{journal}{Eur.\ Phys.\ J.\ B} \textbf{\bibinfo{volume}{35}},
  \bibinfo{pages}{311} (\bibinfo{year}{2003}).

\bibitem[{\citenamefont{Lee et~al.}(2004)\citenamefont{Lee, Louca, Ueda, Park,
  Sato, Isobe, Ueda, Rosenkranz, Zschack, Iniguez et~al.}}]{lee04}
\bibinfo{author}{\bibfnamefont{S.~H.} \bibnamefont{Lee}},
  \bibinfo{author}{\bibfnamefont{D.}~\bibnamefont{Louca}},
  \bibinfo{author}{\bibfnamefont{H.}~\bibnamefont{Ueda}},
  \bibinfo{author}{\bibfnamefont{S.}~\bibnamefont{Park}},
  \bibinfo{author}{\bibfnamefont{T.~J.} \bibnamefont{Sato}},
  \bibinfo{author}{\bibfnamefont{M.}~\bibnamefont{Isobe}},
  \bibinfo{author}{\bibfnamefont{Y.}~\bibnamefont{Ueda}},
  \bibinfo{author}{\bibfnamefont{S.}~\bibnamefont{Rosenkranz}},
  \bibinfo{author}{\bibfnamefont{P.}~\bibnamefont{Zschack}},
  \bibinfo{author}{\bibfnamefont{J.}~\bibnamefont{Iniguez}},
  \bibnamefont{et~al.}, \bibinfo{journal}{Phys.\ Rev.\ Lett.}
  \textbf{\bibinfo{volume}{93}}, \bibinfo{pages}{156407}
  (\bibinfo{year}{2004}).

\bibitem[{\citenamefont{Wheeler et~al.}(2010)\citenamefont{Wheeler, Lake,
  Islam, Reehuis, Steffens, Guidi, and Hill}}]{wheeler10}
\bibinfo{author}{\bibfnamefont{E.~M.} \bibnamefont{Wheeler}},
  \bibinfo{author}{\bibfnamefont{B.}~\bibnamefont{Lake}},
  \bibinfo{author}{\bibfnamefont{A.~T. M.~Nazmul} \bibnamefont{Islam}},
  \bibinfo{author}{\bibfnamefont{M.}~\bibnamefont{Reehuis}},
  \bibinfo{author}{\bibfnamefont{P.}~\bibnamefont{Steffens}},
  \bibinfo{author}{\bibfnamefont{T.}~\bibnamefont{Guidi}}, \bibnamefont{and}
  \bibinfo{author}{\bibfnamefont{A.~H.} \bibnamefont{Hill}},
  \bibinfo{journal}{Phys.\ Rev.\ B} \textbf{\bibinfo{volume}{82}},
  \bibinfo{pages}{140406(R)} (\bibinfo{year}{2010}).

\bibitem[{\citenamefont{Mun et~al.}(2014)\citenamefont{Mun, Chern, Pardo,
  Rivadulla, Sinclair, Zhou, Zapf, and Batista}}]{mun14}
\bibinfo{author}{\bibfnamefont{E.~D.} \bibnamefont{Mun}},
  \bibinfo{author}{\bibfnamefont{G.-W.} \bibnamefont{Chern}},
  \bibinfo{author}{\bibfnamefont{V.}~\bibnamefont{Pardo}},
  \bibinfo{author}{\bibfnamefont{F.}~\bibnamefont{Rivadulla}},
  \bibinfo{author}{\bibfnamefont{R.}~\bibnamefont{Sinclair}},
  \bibinfo{author}{\bibfnamefont{H.~D.} \bibnamefont{Zhou}},
  \bibinfo{author}{\bibfnamefont{V.~S.} \bibnamefont{Zapf}}, \bibnamefont{and}
  \bibinfo{author}{\bibfnamefont{C.~D.} \bibnamefont{Batista}},
  \bibinfo{journal}{Phys.\ Rev.\ Lett.} \textbf{\bibinfo{volume}{112}},
  \bibinfo{pages}{017207} (\bibinfo{year}{2014}).

\bibitem[{\citenamefont{Adachi et~al.}(2005)\citenamefont{Adachi, Suzuki, Kato,
  Osaka, Takata, and Katsufuji}}]{adachi05}
\bibinfo{author}{\bibfnamefont{K.}~\bibnamefont{Adachi}},
  \bibinfo{author}{\bibfnamefont{T.}~\bibnamefont{Suzuki}},
  \bibinfo{author}{\bibfnamefont{K.}~\bibnamefont{Kato}},
  \bibinfo{author}{\bibfnamefont{K.}~\bibnamefont{Osaka}},
  \bibinfo{author}{\bibfnamefont{M.}~\bibnamefont{Takata}}, \bibnamefont{and}
  \bibinfo{author}{\bibfnamefont{T.}~\bibnamefont{Katsufuji}},
  \bibinfo{journal}{Phys.\ Rev.\ Lett.} \textbf{\bibinfo{volume}{95}},
  \bibinfo{pages}{197202} (\bibinfo{year}{2005}).

\bibitem[{\citenamefont{Suzuki et~al.}(2007)\citenamefont{Suzuki, Katsumura,
  Taniguchi, Arima, and Katsufuji}}]{suzuki07_2}
\bibinfo{author}{\bibfnamefont{T.}~\bibnamefont{Suzuki}},
  \bibinfo{author}{\bibfnamefont{M.}~\bibnamefont{Katsumura}},
  \bibinfo{author}{\bibfnamefont{K.}~\bibnamefont{Taniguchi}},
  \bibinfo{author}{\bibfnamefont{T.}~\bibnamefont{Arima}}, \bibnamefont{and}
  \bibinfo{author}{\bibfnamefont{T.}~\bibnamefont{Katsufuji}},
  \bibinfo{journal}{Phys.\ Rev.\ Lett.} \textbf{\bibinfo{volume}{98}},
  \bibinfo{pages}{127203} (\bibinfo{year}{2007}).

\bibitem[{\citenamefont{Zhou et~al.}(2007)\citenamefont{Zhou, Lu, and
  Wiebe}}]{zhou07}
\bibinfo{author}{\bibfnamefont{H.~D.} \bibnamefont{Zhou}},
  \bibinfo{author}{\bibfnamefont{J.}~\bibnamefont{Lu}}, \bibnamefont{and}
  \bibinfo{author}{\bibfnamefont{C.~R.} \bibnamefont{Wiebe}},
  \bibinfo{journal}{Phys.\ Rev.\ B} \textbf{\bibinfo{volume}{76}},
  \bibinfo{pages}{174403} (\bibinfo{year}{2007}).

\bibitem[{\citenamefont{Garlea et~al.}(2008)\citenamefont{Garlea, Jin, Mandrus,
  Roessli, Huang, Miller, Schultz, and Nagler}}]{garlea08}
\bibinfo{author}{\bibfnamefont{V.~O.} \bibnamefont{Garlea}},
  \bibinfo{author}{\bibfnamefont{R.}~\bibnamefont{Jin}},
  \bibinfo{author}{\bibfnamefont{D.}~\bibnamefont{Mandrus}},
  \bibinfo{author}{\bibfnamefont{B.}~\bibnamefont{Roessli}},
  \bibinfo{author}{\bibfnamefont{Q.}~\bibnamefont{Huang}},
  \bibinfo{author}{\bibfnamefont{M.}~\bibnamefont{Miller}},
  \bibinfo{author}{\bibfnamefont{A.~J.} \bibnamefont{Schultz}},
  \bibnamefont{and} \bibinfo{author}{\bibfnamefont{S.~E.}
  \bibnamefont{Nagler}}, \bibinfo{journal}{Phys.\ Rev.\ Lett.}
  \textbf{\bibinfo{volume}{100}}, \bibinfo{pages}{066404}
  (\bibinfo{year}{2008}).

\bibitem[{\citenamefont{Chung et~al.}(2008)\citenamefont{Chung, Kim, Lee, Sato,
  Suzuki, Katsumura, and Katsufuji}}]{chung08}
\bibinfo{author}{\bibfnamefont{J.~H.} \bibnamefont{Chung}},
  \bibinfo{author}{\bibfnamefont{J.~H.} \bibnamefont{Kim}},
  \bibinfo{author}{\bibfnamefont{S.~H.} \bibnamefont{Lee}},
  \bibinfo{author}{\bibfnamefont{T.~J.} \bibnamefont{Sato}},
  \bibinfo{author}{\bibfnamefont{T.}~\bibnamefont{Suzuki}},
  \bibinfo{author}{\bibfnamefont{M.}~\bibnamefont{Katsumura}},
  \bibnamefont{and}
  \bibinfo{author}{\bibfnamefont{T.}~\bibnamefont{Katsufuji}},
  \bibinfo{journal}{Phys.\ Rev.\ B} \textbf{\bibinfo{volume}{77}},
  \bibinfo{pages}{054412} (\bibinfo{year}{2008}).

\bibitem[{\citenamefont{Nii et~al.}(2013)\citenamefont{Nii, Abe, and hisa
  Arima}}]{nii13}
\bibinfo{author}{\bibfnamefont{Y.}~\bibnamefont{Nii}},
  \bibinfo{author}{\bibfnamefont{N.}~\bibnamefont{Abe}}, \bibnamefont{and}
  \bibinfo{author}{\bibfnamefont{T.H.}~\bibnamefont{Arima}},
  \bibinfo{journal}{Phys.\ Rev.\ B} \textbf{\bibinfo{volume}{87}},
  \bibinfo{pages}{085111} (\bibinfo{year}{2013}).

\bibitem[{\citenamefont{Gleason et~al.}(2014)\citenamefont{Gleason, Byrum, Gim,
  Thaler, Abbamonte, MacDougall, Martin, Zhou, and Cooper}}]{gleason14}
\bibinfo{author}{\bibfnamefont{S.~L.} \bibnamefont{Gleason}},
  \bibinfo{author}{\bibfnamefont{T.}~\bibnamefont{Byrum}},
  \bibinfo{author}{\bibfnamefont{Y.}~\bibnamefont{Gim}},
  \bibinfo{author}{\bibfnamefont{A.}~\bibnamefont{Thaler}},
  \bibinfo{author}{\bibfnamefont{P.}~\bibnamefont{Abbamonte}},
  \bibinfo{author}{\bibfnamefont{G.~J.} \bibnamefont{MacDougall}},
  \bibinfo{author}{\bibfnamefont{L.~W.} \bibnamefont{Martin}},
  \bibinfo{author}{\bibfnamefont{H.~D.} \bibnamefont{Zhou}}, \bibnamefont{and}
  \bibinfo{author}{\bibfnamefont{S.~L.} \bibnamefont{Cooper}},
  \bibinfo{journal}{Phys.\ Rev.\ B} \textbf{\bibinfo{volume}{89}},
  \bibinfo{pages}{134402} (\bibinfo{year}{2014}).

\bibitem[{\citenamefont{Nii et~al.}(2012)\citenamefont{Nii, Sagayama, Arima,
  Aoyagi, Sakai, Maki, Nishibori, Sawa, Sugimoto, Ohsumi et~al.}}]{nii12}
\bibinfo{author}{\bibfnamefont{Y.}~\bibnamefont{Nii}},
  \bibinfo{author}{\bibfnamefont{H.}~\bibnamefont{Sagayama}},
  \bibinfo{author}{\bibfnamefont{T.}~\bibnamefont{Arima}},
  \bibinfo{author}{\bibfnamefont{S.}~\bibnamefont{Aoyagi}},
  \bibinfo{author}{\bibfnamefont{R.}~\bibnamefont{Sakai}},
  \bibinfo{author}{\bibfnamefont{S.}~\bibnamefont{Maki}},
  \bibinfo{author}{\bibfnamefont{E.}~\bibnamefont{Nishibori}},
  \bibinfo{author}{\bibfnamefont{H.}~\bibnamefont{Sawa}},
  \bibinfo{author}{\bibfnamefont{K.}~\bibnamefont{Sugimoto}},
  \bibinfo{author}{\bibfnamefont{H.}~\bibnamefont{Ohsumi}},
  \bibnamefont{et~al.}, \bibinfo{journal}{Phys.\ Rev.\ B}
  \textbf{\bibinfo{volume}{86}}, \bibinfo{pages}{125142}
  (\bibinfo{year}{2012}).

\bibitem[{\citenamefont{Katsufuji et~al.}(2008)\citenamefont{Katsufuji, Suzuki,
  Takei, Shingu, Kato, Osaka, Takata, Sagayama, and Arima}}]{katsufuji08}
\bibinfo{author}{\bibfnamefont{T.}~\bibnamefont{Katsufuji}},
  \bibinfo{author}{\bibfnamefont{T.}~\bibnamefont{Suzuki}},
  \bibinfo{author}{\bibfnamefont{H.}~\bibnamefont{Takei}},
  \bibinfo{author}{\bibfnamefont{M.}~\bibnamefont{Shingu}},
  \bibinfo{author}{\bibfnamefont{K.}~\bibnamefont{Kato}},
  \bibinfo{author}{\bibfnamefont{K.}~\bibnamefont{Osaka}},
  \bibinfo{author}{\bibfnamefont{M.}~\bibnamefont{Takata}},
  \bibinfo{author}{\bibfnamefont{H.}~\bibnamefont{Sagayama}}, \bibnamefont{and}
  \bibinfo{author}{\bibfnamefont{T.}~\bibnamefont{Arima}},
  \bibinfo{journal}{J.\ Phys.\ Soc.\ Japan} \textbf{\bibinfo{volume}{77}},
  \bibinfo{pages}{053708} (\bibinfo{year}{2008}).

\bibitem[{\citenamefont{MacDougall et~al.}(2012)\citenamefont{MacDougall,
  Garlea, Aczel, Zhou, and Nagler}}]{macdougall12}
\bibinfo{author}{\bibfnamefont{G.~J.} \bibnamefont{MacDougall}},
  \bibinfo{author}{\bibfnamefont{V.~O.} \bibnamefont{Garlea}},
  \bibinfo{author}{\bibfnamefont{A.~A.} \bibnamefont{Aczel}},
  \bibinfo{author}{\bibfnamefont{H.~D.} \bibnamefont{Zhou}}, \bibnamefont{and}
  \bibinfo{author}{\bibfnamefont{S.~E.} \bibnamefont{Nagler}},
  \bibinfo{journal}{Phys.\ Rev.\ B} \textbf{\bibinfo{volume}{86}},
  \bibinfo{pages}{060414(R)} (\bibinfo{year}{2012}).

\bibitem[{\citenamefont{Zhang et~al.}(2012)\citenamefont{Zhang, Singh, Guillou,
  Simon, Breard, Caignaert, and Hardy}}]{zhang12}
\bibinfo{author}{\bibfnamefont{Q.}~\bibnamefont{Zhang}},
  \bibinfo{author}{\bibfnamefont{K.}~\bibnamefont{Singh}},
  \bibinfo{author}{\bibfnamefont{F.}~\bibnamefont{Guillou}},
  \bibinfo{author}{\bibfnamefont{C.}~\bibnamefont{Simon}},
  \bibinfo{author}{\bibfnamefont{Y.}~\bibnamefont{Breard}},
  \bibinfo{author}{\bibfnamefont{V.}~\bibnamefont{Caignaert}},
  \bibnamefont{and} \bibinfo{author}{\bibfnamefont{V.}~\bibnamefont{Hardy}},
  \bibinfo{journal}{Phys.\ Rev.\ B} \textbf{\bibinfo{volume}{85}},
  \bibinfo{pages}{054405} (\bibinfo{year}{2012}).

\bibitem[{\citenamefont{Kang et~al.}(2012)\citenamefont{Kang, Hwang, Kim, Lee,
  Kim, Kim, Kwon, Lee, Kim, Ueno et~al.}}]{kang12}
\bibinfo{author}{\bibfnamefont{J.-S.} \bibnamefont{Kang}},
  \bibinfo{author}{\bibfnamefont{J.}~\bibnamefont{Hwang}},
  \bibinfo{author}{\bibfnamefont{D.~H.} \bibnamefont{Kim}},
  \bibinfo{author}{\bibfnamefont{E.}~\bibnamefont{Lee}},
  \bibinfo{author}{\bibfnamefont{W.~C.} \bibnamefont{Kim}},
  \bibinfo{author}{\bibfnamefont{C.~S.} \bibnamefont{Kim}},
  \bibinfo{author}{\bibfnamefont{S.}~\bibnamefont{Kwon}},
  \bibinfo{author}{\bibfnamefont{S.}~\bibnamefont{Lee}},
  \bibinfo{author}{\bibfnamefont{J.-Y.} \bibnamefont{Kim}},
  \bibinfo{author}{\bibfnamefont{T.}~\bibnamefont{Ueno}}, \bibnamefont{et~al.},
  \bibinfo{journal}{Phys.\ Rev.\ B} \textbf{\bibinfo{volume}{85}},
  \bibinfo{pages}{165136} (\bibinfo{year}{2012}).

\bibitem[{\citenamefont{MacDougall et~al.}(2014)\citenamefont{MacDougall,
  Brodsky, Aczel, Garlea, Granroth, Christianson, Hong, Zhou, and
  Nagler}}]{macdougall14}
\bibinfo{author}{\bibfnamefont{G.~J.} \bibnamefont{MacDougall}},
  \bibinfo{author}{\bibfnamefont{I.}~\bibnamefont{Brodsky}},
  \bibinfo{author}{\bibfnamefont{A.~A.} \bibnamefont{Aczel}},
  \bibinfo{author}{\bibfnamefont{V.~O.} \bibnamefont{Garlea}},
  \bibinfo{author}{\bibfnamefont{G.~E.} \bibnamefont{Granroth}},
  \bibinfo{author}{\bibfnamefont{A.~D.} \bibnamefont{Christianson}},
  \bibinfo{author}{\bibfnamefont{T.}~\bibnamefont{Hong}},
  \bibinfo{author}{\bibfnamefont{H.~D.} \bibnamefont{Zhou}}, \bibnamefont{and}
  \bibinfo{author}{\bibfnamefont{S.~E.} \bibnamefont{Nagler}},
  \bibinfo{journal}{Phys.\ Rev.\ B} \textbf{\bibinfo{volume}{89}},
  \bibinfo{pages}{224404} (\bibinfo{year}{2014}).

\bibitem[{\citenamefont{Zhang et~al.}(2014)\citenamefont{Zhang, Ramazanoglu,
  Chi, Liu, Lograsso, and Vaknin}}]{zhang14}
\bibinfo{author}{\bibfnamefont{Q.}~\bibnamefont{Zhang}},
  \bibinfo{author}{\bibfnamefont{M.}~\bibnamefont{Ramazanoglu}},
  \bibinfo{author}{\bibfnamefont{S.}~\bibnamefont{Chi}},
  \bibinfo{author}{\bibfnamefont{Y.}~\bibnamefont{Liu}},
  \bibinfo{author}{\bibfnamefont{T.~A.} \bibnamefont{Lograsso}},
  \bibnamefont{and} \bibinfo{author}{\bibfnamefont{D.}~\bibnamefont{Vaknin}},
  \bibinfo{journal}{Phys.\ Rev.\ B} \textbf{\bibinfo{volume}{89}},
  \bibinfo{pages}{224416} (\bibinfo{year}{2014}).

\bibitem[{\citenamefont{Menyuk et~al.}(1962)\citenamefont{Menyuk, Wold, Rogers,
  and Dwight}}]{menyuk62}
\bibinfo{author}{\bibfnamefont{N.}~\bibnamefont{Menyuk}},
  \bibinfo{author}{\bibfnamefont{A.}~\bibnamefont{Wold}},
  \bibinfo{author}{\bibfnamefont{D.}~\bibnamefont{Rogers}}, \bibnamefont{and}
  \bibinfo{author}{\bibfnamefont{K.}~\bibnamefont{Dwight}},
  \bibinfo{journal}{Journal of Applied Physics} \textbf{\bibinfo{volume}{33}},
  \bibinfo{pages}{1144} (\bibinfo{year}{1962}).

\bibitem[{\citenamefont{Dwight et~al.}(1965)\citenamefont{Dwight, Menyuk,
  Rogers, and Wold}}]{dwight64}
\bibinfo{author}{\bibfnamefont{K.}~\bibnamefont{Dwight}},
  \bibinfo{author}{\bibfnamefont{N.}~\bibnamefont{Menyuk}},
  \bibinfo{author}{\bibfnamefont{D.~B.} \bibnamefont{Rogers}},
  \bibnamefont{and} \bibinfo{author}{\bibfnamefont{A.}~\bibnamefont{Wold}},
  \bibinfo{journal}{Proc. Int. Conf. on Magnetism 1964 (Nottingham)} pp.
  \bibinfo{pages}{538--541} (\bibinfo{year}{1965}).

\bibitem[{\citenamefont{Kismarahardja et~al.}(2011)\citenamefont{Kismarahardja,
  Brooks, Kiswandhi, Matsubayashi, Yamanaka, Uwatoko, Whalen, Siegrist, and
  Zhou}}]{kismarahardja11}
\bibinfo{author}{\bibfnamefont{A.}~\bibnamefont{Kismarahardja}},
  \bibinfo{author}{\bibfnamefont{J.~S.} \bibnamefont{Brooks}},
  \bibinfo{author}{\bibfnamefont{A.}~\bibnamefont{Kiswandhi}},
  \bibinfo{author}{\bibfnamefont{K.}~\bibnamefont{Matsubayashi}},
  \bibinfo{author}{\bibfnamefont{R.}~\bibnamefont{Yamanaka}},
  \bibinfo{author}{\bibfnamefont{Y.}~\bibnamefont{Uwatoko}},
  \bibinfo{author}{\bibfnamefont{J.}~\bibnamefont{Whalen}},
  \bibinfo{author}{\bibfnamefont{T.}~\bibnamefont{Siegrist}}, \bibnamefont{and}
  \bibinfo{author}{\bibfnamefont{H.~D.} \bibnamefont{Zhou}},
  \bibinfo{journal}{Phys.\ Rev.\ Lett.} \textbf{\bibinfo{volume}{106}},
  \bibinfo{pages}{056602} (\bibinfo{year}{2011}).

\bibitem[{\citenamefont{Huang et~al.}(2012)\citenamefont{Huang, Yang, and
  Zhang}}]{huang12}
\bibinfo{author}{\bibfnamefont{Y.}~\bibnamefont{Huang}},
  \bibinfo{author}{\bibfnamefont{Z.}~\bibnamefont{Yang}}, \bibnamefont{and}
  \bibinfo{author}{\bibfnamefont{Y.}~\bibnamefont{Zhang}},
  \bibinfo{journal}{J.\ Phys.: Condens.\ Matter} \textbf{\bibinfo{volume}{24}},
  \bibinfo{pages}{056003} (\bibinfo{year}{2012}).

\bibitem[{\citenamefont{Koborinai et~al.}(2015)\citenamefont{Koborinai,
  Dissanayake, Reehuis, Matsuda, Lee, and Katsufuji}}]{koborinai15}
\bibinfo{author}{\bibfnamefont{R.}~\bibnamefont{Koborinai}},
  \bibinfo{author}{\bibfnamefont{S.~E.} \bibnamefont{Dissanayake}},
  \bibinfo{author}{\bibfnamefont{M.}~\bibnamefont{Reehuis}},
  \bibinfo{author}{\bibfnamefont{M.}~\bibnamefont{Matsuda}},
  \bibinfo{author}{\bibfnamefont{S.-H.} \bibnamefont{Lee}}, \bibnamefont{and}
  \bibinfo{author}{\bibfnamefont{T.}~\bibnamefont{Katsufuji}},
  \bibinfo{journal}{arXiv:1505.04864v1}  (\bibinfo{year}{2015}).

\bibitem[{\citenamefont{Motome and Tsunetsugu}(2004)}]{tsunetsugu03}
\bibinfo{author}{\bibfnamefont{Y.}~\bibnamefont{Motome}} \bibnamefont{and}
  \bibinfo{author}{\bibfnamefont{H.}~\bibnamefont{Tsunetsugu}},
  \bibinfo{journal}{Phys.\ Rev.\ B} \textbf{\bibinfo{volume}{70}},
  \bibinfo{pages}{184427} (\bibinfo{year}{2004}).

\bibitem[{\citenamefont{Tsunetsugu and Motome}(2003)}]{motome04}
 \bibinfo{author}{\bibfnamefont{H.}~\bibnamefont{Tsunetsugu}}\bibnamefont{and}
\bibinfo{author}{\bibfnamefont{Y.}~\bibnamefont{Motome}},
  \bibinfo{journal}{Phys.\ Rev.\ B} \textbf{\bibinfo{volume}{68}},
  \bibinfo{pages}{060405(R)} (\bibinfo{year}{2003}).

\bibitem[{\citenamefont{Tchernyshyov}(2004)}]{tchernyshyov04}
\bibinfo{author}{\bibfnamefont{O.}~\bibnamefont{Tchernyshyov}},
  \bibinfo{journal}{Phys.\ Rev.\ Lett.} \textbf{\bibinfo{volume}{93}},
  \bibinfo{pages}{157206} (\bibinfo{year}{2004}).

\bibitem[{\citenamefont{Maitra and Valenti}(2007)}]{maitra07}
\bibinfo{author}{\bibfnamefont{T.}~\bibnamefont{Maitra}} \bibnamefont{and}
  \bibinfo{author}{\bibfnamefont{R.}~\bibnamefont{Valenti}},
  \bibinfo{journal}{Phys.\ Rev.\ Lett.} \textbf{\bibinfo{volume}{99}},
  \bibinfo{pages}{126401} (\bibinfo{year}{2007}).

\bibitem[{\citenamefont{Perkins and Sikora}(2007)}]{perkins07}
\bibinfo{author}{\bibfnamefont{N.~B.} \bibnamefont{Perkins}} \bibnamefont{and}
  \bibinfo{author}{\bibfnamefont{O.}~\bibnamefont{Sikora}},
  \bibinfo{journal}{Phys.\ Rev.\ B} \textbf{\bibinfo{volume}{76}},
  \bibinfo{pages}{214434} (\bibinfo{year}{2007}).

\bibitem[{\citenamefont{Sarkar et~al.}(2009)\citenamefont{Sarkar, Maitra,
  Valenti, and Saha-Dasgupta}}]{sarkar09}
\bibinfo{author}{\bibfnamefont{S.}~\bibnamefont{Sarkar}},
  \bibinfo{author}{\bibfnamefont{T.}~\bibnamefont{Maitra}},
  \bibinfo{author}{\bibfnamefont{R.}~\bibnamefont{Valenti}}, \bibnamefont{and}
  \bibinfo{author}{\bibfnamefont{T.}~\bibnamefont{Saha-Dasgupta}},
  \bibinfo{journal}{Phys.\ Rev.\ Lett.} \textbf{\bibinfo{volume}{102}},
  \bibinfo{pages}{216405} (\bibinfo{year}{2009}).

\bibitem[{\citenamefont{Chern et~al.}(2010)\citenamefont{Chern, Perkins, and
  Hao}}]{chern10}
\bibinfo{author}{\bibfnamefont{G.~W.} \bibnamefont{Chern}},
  \bibinfo{author}{\bibfnamefont{N.}~\bibnamefont{Perkins}}, \bibnamefont{and}
  \bibinfo{author}{\bibfnamefont{Z.}~\bibnamefont{Hao}},
  \bibinfo{journal}{Phys.\ Rev.\ B} \textbf{\bibinfo{volume}{81}},
  \bibinfo{pages}{125127} (\bibinfo{year}{2010}).

\bibitem[{\citenamefont{Sarkar and Saha-Dasgupta}(2011)}]{sarkar11}
\bibinfo{author}{\bibfnamefont{S.}~\bibnamefont{Sarkar}} \bibnamefont{and}
  \bibinfo{author}{\bibfnamefont{T.}~\bibnamefont{Saha-Dasgupta}},
  \bibinfo{journal}{Phys.\ Rev.\ B} \textbf{\bibinfo{volume}{84}},
  \bibinfo{pages}{235112} (\bibinfo{year}{2011}).

\bibitem[{\citenamefont{Rogers et~al.}(1963)\citenamefont{Rogers, Arnott, Wold,
  and Goodenough}}]{rogers63}
\bibinfo{author}{\bibfnamefont{D.~B.} \bibnamefont{Rogers}},
  \bibinfo{author}{\bibfnamefont{R.~J.} \bibnamefont{Arnott}},
  \bibinfo{author}{\bibfnamefont{A.}~\bibnamefont{Wold}}, \bibnamefont{and}
  \bibinfo{author}{\bibfnamefont{J.~B.} \bibnamefont{Goodenough}},
  \bibinfo{journal}{J.\ Phys.\ Chem.\ Solids} \textbf{\bibinfo{volume}{24}},
  \bibinfo{pages}{347} (\bibinfo{year}{1963}).

\bibitem[{\citenamefont{Rogers et~al.}(1964)\citenamefont{Rogers, Goodenough,
  and Wold}}]{rogers64}
\bibinfo{author}{\bibfnamefont{D.~B.} \bibnamefont{Rogers}},
  \bibinfo{author}{\bibfnamefont{J.~B.} \bibnamefont{Goodenough}},
  \bibnamefont{and} \bibinfo{author}{\bibfnamefont{A.}~\bibnamefont{Wold}},
  \bibinfo{journal}{J.\ Applied Physics} \textbf{\bibinfo{volume}{35}},
  \bibinfo{pages}{1069} (\bibinfo{year}{1964}).

\bibitem[{\citenamefont{Blanco-Canosa et~al.}(2007)\citenamefont{Blanco-Canosa,
  Rivadulla, Pardo, Baldomir, Zhou, Garcia-Hernandez, Lopez-Quintela, Rivas,
  and Goodenough}}]{blanco07}
\bibinfo{author}{\bibfnamefont{S.}~\bibnamefont{Blanco-Canosa}},
  \bibinfo{author}{\bibfnamefont{F.}~\bibnamefont{Rivadulla}},
  \bibinfo{author}{\bibfnamefont{V.}~\bibnamefont{Pardo}},
  \bibinfo{author}{\bibfnamefont{D.}~\bibnamefont{Baldomir}},
  \bibinfo{author}{\bibfnamefont{J.-S.} \bibnamefont{Zhou}},
  \bibinfo{author}{\bibfnamefont{M.}~\bibnamefont{Garcia-Hernandez}},
  \bibinfo{author}{\bibfnamefont{M.~A.} \bibnamefont{Lopez-Quintela}},
  \bibinfo{author}{\bibfnamefont{J.}~\bibnamefont{Rivas}}, \bibnamefont{and}
  \bibinfo{author}{\bibfnamefont{J.~B.} \bibnamefont{Goodenough}},
  \bibinfo{journal}{Phys.\ Rev.\ Lett.} \textbf{\bibinfo{volume}{99}},
  \bibinfo{pages}{187201} (\bibinfo{year}{2007}).

\bibitem[{\citenamefont{Pardo et~al.}(2008)\citenamefont{Pardo, Blanco-Canosa,
  Rivadulla, Khomskii, Baldomir, Wu, and Rivas}}]{pardo08}
\bibinfo{author}{\bibfnamefont{V.}~\bibnamefont{Pardo}},
  \bibinfo{author}{\bibfnamefont{S.}~\bibnamefont{Blanco-Canosa}},
  \bibinfo{author}{\bibfnamefont{F.}~\bibnamefont{Rivadulla}},
  \bibinfo{author}{\bibfnamefont{D.~I.} \bibnamefont{Khomskii}},
  \bibinfo{author}{\bibfnamefont{D.}~\bibnamefont{Baldomir}},
  \bibinfo{author}{\bibfnamefont{H.}~\bibnamefont{Wu}}, \bibnamefont{and}
  \bibinfo{author}{\bibfnamefont{J.}~\bibnamefont{Rivas}},
  \bibinfo{journal}{Phys.\ Rev.\ Lett.} \textbf{\bibinfo{volume}{101}},
  \bibinfo{pages}{256403} (\bibinfo{year}{2008}).

\bibitem[{\citenamefont{Kaur et~al.}(2014)\citenamefont{Kaur, Maitra, and
  Nautiyal}}]{kaur14}
\bibinfo{author}{\bibfnamefont{R.}~\bibnamefont{Kaur}},
  \bibinfo{author}{\bibfnamefont{T.}~\bibnamefont{Maitra}}, \bibnamefont{and}
  \bibinfo{author}{\bibfnamefont{T.}~\bibnamefont{Nautiyal}},
  \bibinfo{journal}{J.\ Phys.: Condens.\ Matter} \textbf{\bibinfo{volume}{26}},
  \bibinfo{pages}{045505} (\bibinfo{year}{2014}).

\bibitem[{\citenamefont{Kismarahardja et~al.}(2013)\citenamefont{Kismarahardja,
  Brooks, Zhou, Choi, Matsubayashi, and Uwatoko}}]{kismarahardja13}
\bibinfo{author}{\bibfnamefont{A.}~\bibnamefont{Kismarahardja}},
  \bibinfo{author}{\bibfnamefont{J.~S.} \bibnamefont{Brooks}},
  \bibinfo{author}{\bibfnamefont{H.~D.} \bibnamefont{Zhou}},
  \bibinfo{author}{\bibfnamefont{E.~S.} \bibnamefont{Choi}},
  \bibinfo{author}{\bibfnamefont{K.}~\bibnamefont{Matsubayashi}},
  \bibnamefont{and} \bibinfo{author}{\bibfnamefont{Y.}~\bibnamefont{Uwatoko}},
  \bibinfo{journal}{Phys.\ Rev.\ B} \textbf{\bibinfo{volume}{87}},
  \bibinfo{pages}{054432} (\bibinfo{year}{2013}).

\bibitem[{\citenamefont{Kiswandhi et~al.}(2011)\citenamefont{Kiswandhi, Brooks,
  Lu, Whalen, Siegrist, and Zhou}}]{kiswandhi11}
\bibinfo{author}{\bibfnamefont{A.}~\bibnamefont{Kiswandhi}},
  \bibinfo{author}{\bibfnamefont{J.~S.} \bibnamefont{Brooks}},
  \bibinfo{author}{\bibfnamefont{J.}~\bibnamefont{Lu}},
  \bibinfo{author}{\bibfnamefont{J.}~\bibnamefont{Whalen}},
  \bibinfo{author}{\bibfnamefont{T.}~\bibnamefont{Siegrist}}, \bibnamefont{and}
  \bibinfo{author}{\bibfnamefont{H.~D.} \bibnamefont{Zhou}},
  \bibinfo{journal}{Phys.\ Rev.\ B} \textbf{\bibinfo{volume}{84}},
  \bibinfo{pages}{205138} (\bibinfo{year}{2011}).

\bibitem[{\citenamefont{Ma et~al.}(2015)\citenamefont{Ma, Lee, Hahn, aand
  H.~B.~Cao, Aczel, Dun, Stone, Tian, Qiu, Copley et~al.}}]{ma15}
\bibinfo{author}{\bibfnamefont{J.}~\bibnamefont{Ma}},
  \bibinfo{author}{\bibfnamefont{J.~H.} \bibnamefont{Lee}},
  \bibinfo{author}{\bibfnamefont{S.~E.} \bibnamefont{Hahn}},
  \bibinfo{author}{\bibfnamefont{T.~H.} \bibnamefont{aand H.~B.~Cao}},
  \bibinfo{author}{\bibfnamefont{A.~A.} \bibnamefont{Aczel}},
  \bibinfo{author}{\bibfnamefont{Z.~L.} \bibnamefont{Dun}},
  \bibinfo{author}{\bibfnamefont{M.~B.} \bibnamefont{Stone}},
  \bibinfo{author}{\bibfnamefont{W.}~\bibnamefont{Tian}},
  \bibinfo{author}{\bibfnamefont{Q.}~\bibnamefont{Qiu}},
  \bibinfo{author}{\bibfnamefont{J.~R.~D.} \bibnamefont{Copley}},
  \bibnamefont{et~al.}, \bibinfo{journal}{Phys.\ Rev.\ B}
  \textbf{\bibinfo{volume}{91}}, \bibinfo{pages}{020407(R)}
  (\bibinfo{year}{2015}).

\bibitem[{\citenamefont{Khomskii and Mizokawa}(2005)}]{khomskii05}
\bibinfo{author}{\bibfnamefont{D.~I.} \bibnamefont{Khomskii}} \bibnamefont{and}
  \bibinfo{author}{\bibfnamefont{T.}~\bibnamefont{Mizokawa}},
  \bibinfo{journal}{Phys.\ Rev.\ Lett.} \textbf{\bibinfo{volume}{94}},
  \bibinfo{pages}{156402} (\bibinfo{year}{2005}).

\bibitem[{\citenamefont{Fichtl et~al.}(2005)\citenamefont{Fichtl, Tsurkan,
  Lunkenheimer, Hemberger, Fritsch, von Nidda, Scheidt, and Loidl}}]{fichtl05}
\bibinfo{author}{\bibfnamefont{R.}~\bibnamefont{Fichtl}},
  \bibinfo{author}{\bibfnamefont{V.}~\bibnamefont{Tsurkan}},
  \bibinfo{author}{\bibfnamefont{P.}~\bibnamefont{Lunkenheimer}},
  \bibinfo{author}{\bibfnamefont{J.}~\bibnamefont{Hemberger}},
  \bibinfo{author}{\bibfnamefont{V.}~\bibnamefont{Fritsch}},
  \bibinfo{author}{\bibfnamefont{H.-A.} \bibnamefont{Krug von Nidda}},
  \bibinfo{author}{\bibfnamefont{E.-W.} \bibnamefont{Scheidt}},
  \bibnamefont{and} \bibinfo{author}{\bibfnamefont{A.}~\bibnamefont{Loidl}},
  \bibinfo{journal}{Phys.\ Rev.\ Lett.} \textbf{\bibinfo{volume}{94}},
  \bibinfo{pages}{027601} (\bibinfo{year}{2005}).

\bibitem[{\citenamefont{MacDougall et~al.}(2011)\citenamefont{MacDougall, Gout,
  Zarestky, Ehlers, Podlesnyak, McGuire, Mandrus, and Nagler}}]{macdougall11}
\bibinfo{author}{\bibfnamefont{G.~J.} \bibnamefont{MacDougall}},
  \bibinfo{author}{\bibfnamefont{D.}~\bibnamefont{Gout}},
  \bibinfo{author}{\bibfnamefont{J.~L.} \bibnamefont{Zarestky}},
  \bibinfo{author}{\bibfnamefont{G.}~\bibnamefont{Ehlers}},
  \bibinfo{author}{\bibfnamefont{A.}~\bibnamefont{Podlesnyak}},
  \bibinfo{author}{\bibfnamefont{M.~A.} \bibnamefont{McGuire}},
  \bibinfo{author}{\bibfnamefont{D.}~\bibnamefont{Mandrus}}, \bibnamefont{and}
  \bibinfo{author}{\bibfnamefont{S.~E.} \bibnamefont{Nagler}},
  \bibinfo{journal}{Proc.\ Nat.\ Acad.\ Sci.} \textbf{\bibinfo{volume}{108}},
  \bibinfo{pages}{15693-15698}
  (\bibinfo{year}{2011}).

\bibitem[{\citenamefont{Rogers et~al.}(1966)\citenamefont{Rogers, Ferretti, and
  Kunnmann}}]{rogers66}
\bibinfo{author}{\bibfnamefont{D.~B.} \bibnamefont{Rogers}},
  \bibinfo{author}{\bibfnamefont{A.}~\bibnamefont{Ferretti}}, \bibnamefont{and}
  \bibinfo{author}{\bibfnamefont{W.}~\bibnamefont{Kunnmann}},
  \bibinfo{journal}{J.\ Phys.\ Chem.\ Solids} \textbf{\bibinfo{volume}{27}},
  \bibinfo{pages}{1445} (\bibinfo{year}{1966}).

\bibitem[{\citenamefont{Garlea et~al.}(2010)\citenamefont{Garlea, Chakoumakos,
  Moore, Taylor, Chae, Maples, Riedel, Lynn, and Selby}}]{hb2a}
\bibinfo{author}{\bibfnamefont{V.}~\bibnamefont{Garlea}},
  \bibinfo{author}{\bibfnamefont{B.}~\bibnamefont{Chakoumakos}},
  \bibinfo{author}{\bibfnamefont{S.}~\bibnamefont{Moore}},
  \bibinfo{author}{\bibfnamefont{G.}~\bibnamefont{Taylor}},
  \bibinfo{author}{\bibfnamefont{T.}~\bibnamefont{Chae}},
  \bibinfo{author}{\bibfnamefont{R.}~\bibnamefont{Maples}},
  \bibinfo{author}{\bibfnamefont{R.}~\bibnamefont{Riedel}},
  \bibinfo{author}{\bibfnamefont{G.}~\bibnamefont{Lynn}}, \bibnamefont{and}
  \bibinfo{author}{\bibfnamefont{D.}~\bibnamefont{Selby}},
  \bibinfo{journal}{Appl.\ Phys.\ A} \textbf{\bibinfo{volume}{99}},
  \bibinfo{pages}{531} (\bibinfo{year}{2010}).

\bibitem[{\citenamefont{Neuefeind et~al.}(2012)\citenamefont{Neuefeind,
  Feygenson, Carruth, Hoffmann, and Chipley}}]{nomad}
\bibinfo{author}{\bibfnamefont{J.}~\bibnamefont{Neuefeind}},
  \bibinfo{author}{\bibfnamefont{M.}~\bibnamefont{Feygenson}},
  \bibinfo{author}{\bibfnamefont{J.}~\bibnamefont{Carruth}},
  \bibinfo{author}{\bibfnamefont{R.}~\bibnamefont{Hoffmann}}, \bibnamefont{and}
  \bibinfo{author}{\bibfnamefont{K.}~\bibnamefont{Chipley}},
  \bibinfo{journal}{Nuclear Instruments and Methods B}
  \textbf{\bibinfo{volume}{287}}, \bibinfo{pages}{68} (\bibinfo{year}{2012}).


\bibitem[{\citenamefont{Rodriguez-Carvajal}(2012)\citenamefont{Rodriguez-Carvajal}}]{fullprof}
\bibinfo{author}{\bibfnamefont{J.}~\bibnamefont{Rodriguez-Carvajal}},
  \bibinfo{journal}{Physica B}
  \textbf{\bibinfo{volume}{192}}, \bibinfo{pages}{55} (\bibinfo{year}{1993}).


\bibitem[{\citenamefont{Farrow et~al.}(2012)\citenamefont{Farrow,
  Juhas, Liu, Bryndin, Bozin, Block, Proffen and Billinge}}]{pdfgui}
\bibinfo{author}{\bibfnamefont{C. L.}~\bibnamefont{Farrow}},
  \bibinfo{author}{\bibfnamefont{P.}~\bibnamefont{Juhas}},
  \bibinfo{author}{\bibfnamefont{J. W.}~\bibnamefont{Liu}},
  \bibinfo{author}{\bibfnamefont{D.}~\bibnamefont{Bryndin}},
   \bibinfo{author}{\bibfnamefont{E. S.}~\bibnamefont{Bozin}},
    \bibinfo{author}{\bibfnamefont{J.}~\bibnamefont{Bloch}},
   \bibinfo{author}{\bibfnamefont{Th.}~\bibnamefont{Proffen}},
  \bibnamefont{and}
  \bibinfo{author}{\bibfnamefont{S. J. L.}~\bibnamefont{Billinge}},
  \bibinfo{journal}{J. Phys: Condens. Mat.}
  \textbf{\bibinfo{volume}{19}}, \bibinfo{pages}{335219} (\bibinfo{year}{2007}).


\bibitem[{\citenamefont{Winn et~al.}(2015)\citenamefont{Winn, Filges, Garlea,
  Graves-Brook, Hagen, Jiang, Kenzelmann, Passell, Shapiro, Tong
  et~al.}}]{hyspec}
\bibinfo{author}{\bibfnamefont{B.}~\bibnamefont{Winn}},
  \bibinfo{author}{\bibfnamefont{U.}~\bibnamefont{Filges}},
  \bibinfo{author}{\bibfnamefont{V.~O.} \bibnamefont{Garlea}},
  \bibinfo{author}{\bibfnamefont{M.}~\bibnamefont{Graves-Brook}},
  \bibinfo{author}{\bibfnamefont{M.}~\bibnamefont{Hagen}},
  \bibinfo{author}{\bibfnamefont{C.}~\bibnamefont{Jiang}},
  \bibinfo{author}{\bibfnamefont{M.}~\bibnamefont{Kenzelmann}},
  \bibinfo{author}{\bibfnamefont{L.}~\bibnamefont{Passell}},
  \bibinfo{author}{\bibfnamefont{S.~M.} \bibnamefont{Shapiro}},
  \bibinfo{author}{\bibfnamefont{X.}~\bibnamefont{Tong}}, \bibnamefont{et~al.},
  \bibinfo{journal}{EPJ Web of Conferences} \textbf{\bibinfo{volume}{83}},
  \bibinfo{pages}{03017} (\bibinfo{year}{2015}).

\bibitem[{\citenamefont{McWhan et~al.}(1973)\citenamefont{McWhan, Menth,
  Remeika, Brinkman, and Rice}}]{mcwhan73}
\bibinfo{author}{\bibfnamefont{D.~B.} \bibnamefont{McWhan}},
  \bibinfo{author}{\bibfnamefont{A.}~\bibnamefont{Menth}},
  \bibinfo{author}{\bibfnamefont{J.~P.} \bibnamefont{Remeika}},
  \bibinfo{author}{\bibfnamefont{W.~F.} \bibnamefont{Brinkman}},
  \bibnamefont{and} \bibinfo{author}{\bibfnamefont{T.~M.} \bibnamefont{Rice}},
  \bibinfo{journal}{Phys.\ Rev.\ B} \textbf{\bibinfo{volume}{7}},
  \bibinfo{pages}{1920} (\bibinfo{year}{1973}).

\end{thebibliography}

\end{document}